%% file: main.tex
\documentclass[useAMS,usenatbib,times,letter,amssymb]{mnras}
\usepackage{epsfig,times,amssymb,amsmath,verbatim,xspace}
\usepackage[usenames,dvipsnames,svgnames,table]{xcolor}
\usepackage{tablefootnote}
\usepackage{todonotes}
\usepackage{tabularx}
\usepackage{float}
\usepackage{fancyhdr}
\usepackage{natbib,hyperref,ifthen,soul} 
\usepackage{eso-pic}

\AddToShipoutPictureBG*{
  \AtPageUpperLeft{
    \hspace{0.75\paperwidth}
    \raisebox{-9.\baselineskip}{
      \makebox[0pt][l]{\textnormal{DES 2017-0727}}
}}}

\AddToShipoutPictureBG*{
  \AtPageUpperLeft{
    \hspace{0.75\paperwidth}
    \raisebox{-10.\baselineskip}{
      \makebox[0pt][l]{\textnormal{FERMILAB-PUB-17-291-E}}
}}}

\title[Cosmic Shear \& Galaxy Neighbours]{Dark Energy Survey Year 1 Results: The Impact of Galaxy Neighbours on Weak Lensing Cosmology with {\sc im3shape}}
\input{authors}

\defcitealias{jarvis16}{J16}
\defcitealias{shearcat}{Z17}

\newcommand{\imshape}{{\textsc{im3shape}}}
\newcommand{\metacal}{\textsc{metacalibration}}

\newcommand{\se}{\textsc{SEx\-tractor}}

\newcommand{\hoopoe}{\textsc{Hoopoe}}
\newcommand{\waxwing}{\textsc{Waxwing}}

\newcommand{\sersic}{S\'ersic}
\newcommand{\blockfont}[1]{{\textsc{#1}}\xspace}

\newcommand{\Omegam}{\ensuremath{\Omega_\mathrm{m}}}
\newcommand{\as}{$A_\mathrm{s}$}
\newcommand{\ns}{$n_\mathrm{s}$}

\newcommand{\snr}{\ensuremath{S/N}}
\newcommand{\rgp}{\ensuremath{R_{gp}/R_p}}

\begin{document}
\thispagestyle{empty}
\maketitle

\begin{abstract}
We use a suite of simulated images based on Year 1 of the Dark Energy Survey to explore the impact of galaxy neighbours on shape measurement and shear cosmology. 
The \hoopoe~image simulations include realistic blending, galaxy positions, and spatial variations in depth and PSF properties.
Using the \imshape~maximum-likelihood shape measurement code, we identify four mechanisms by which neighbours can have a non-negligible influence on shear estimation. 
These effects, if ignored, would contribute a net multiplicative bias of $m \sim 0.03 - 0.09$ 
in the DES Y1 \imshape~catalogue, though the precise impact will be dependent on both the measurement code and the selection cuts applied.  
This can be reduced to percentage level or less by removing objects with close neighbours,
at a cost to the effective number density of galaxies $n_\mathrm{eff}$ of 30\%. 
We use the cosmological inference pipeline of DES Y1 to explore the cosmological implications of neighbour bias
and show that omitting blending from the calibration simulation for DES Y1 
would bias the inferred clustering amplitude $S_8\equiv \sigma_8 (\Omegam /0.3)^{0.5}$ by $2 \sigma$ towards low values.
Finally, we use the \hoopoe~simulations to test the effect of neighbour-induced spatial correlations
in the multiplicative bias. We find the impact on the recovered $S_8$ of ignoring such correlations to be subdominant to statistical error at the current level of precision.
\end{abstract}



\section{Introduction}

A standard and well tested prediction of General Relativity is that a 
concentration of mass will distort the spacetime around it, 
and thus produce a curious phenomenon called gravitational lensing. 
The most obvious manifestation is about massive galaxy clusters, 
where background galaxies can be elongated into cresent-shaped arcs. 
So-called strong lensing of galaxies was first observed in
the late 1980s and has been confirmed many times since.
A subtler, but from a cosmologist's perspective more powerful, consequence of gravitational lensing
is that background fluctuations in the density of dark matter will induce coherent distortions to photons' paths. 
This effect is known as cosmic shear, and it was first detected by four groups
at around the same time close to two decades ago \citep{bacon00, vanWaerbeke00, kaiser00, wittman00}.
Cosmic shear has the potential to be the single most powerful probe in the toolbox of modern cosmology. 
The spatial correlations due to lensing are a direct imprint of the large scale mass distribution of the Universe.
Thus it allows one to study the total mass of the Universe and the growth of structure within it 
\citep{maoli01,jarvis06, massey07,kilbinger13,heymans13,svcosmology,jee16,hildebrandt16,koehlinger17}, 
or to map out the spatial distribution of dark matter on the sky
(eg \citealt{kaiser94,vanWaerbeke13,chang15}).
As a probe of both structure and geometry, cosmic shear is also attractive as a method for shedding light
on the as yet poorly understood component of the Universe known as dark energy
\citep{albrecht06,weinberg12}.
Alternatively, lensing will allow us to place ever more stringent tests of our theories of gravity 
\citep{simpson15,harnoisDeraps15,brouwer17}.
Is also theoretically very clean, responding directly to the power spectrum of dark matter,
which is affected by baryonic physics only on small scales, and avoids recourse to poorly-understood
phenomenological rules.
Indeed galaxy number density enters only at second order as a weighting of the observed shear
due to the fact that one can only sample the shear field where there are real galaxies \citep{schmidt09}.

Though well modelled theoretically, cosmic shear is technically highly challenging to measure;
as with all these probes it is not without its own sources of systematic error.
It also cannot be reiterated too many times that the shear component of even the most distant galaxy's shape is
subdominant to noise by an order of magnitude. 
Indeed, the ambitions of the current generation of cosmology surveys will 
require sub-percent level uncertainties
(both systematic and statistical)
on what is already a tiny cosmological ellipticity component $g\sim 0.01$.

It was realised early on how significant the task of translating photometric galaxy 
images into unbiased shear measurements would be.
In response came a series of blind shear measurement challenges, 
designed to review, test and compare the best methods available.
The first of these, called STEP1 \citep{step1} grew out of a discussion at the
225th IAU Symposium in 2004.
The exercise was based around a set of simple \blockfont{SkyMaker} simulations \citep{skymaker}, 
which were designed to mimic ground based observations but with analytic galaxies and PSFs and constant shear.
The algorithms at this point represented a first wave of shear measurement codes
and included several moments-based algorithms \citep{kaiser95, kuijken99, rhodes01}, 
some early forward modelling methods \citep{im2shape},
as well as a technique called shapelets, 
which models a light profile as a set of 2D basis functions \citep{bernstein02,refregier03}.

The simulations and the codes themselves steadily grew in complexity.
STEP2 was followed by series of GREAT challenges 
\citep{step2,great08,great10handbook,mandelbaum14},
which focused on different aspects of shape measurement bias and 
have been essential in quantifying a number of significant effects.
In recent years the drive to find ever more accurate ways to measure shear has intensified,
with many novel approaches being suggested.
For example \citet{fc16} use a form of self-calibration, which repeats the shape measurement 
on a test image based on the best-fitting model for each galaxy. 
A related approach, named metacalibration, involves deriving corrections to the galaxy shape 
measurements directly from the data, using modified copies of the image with additional 
shear \citep{huff17, sheldon17}.
More advanced moments-based approaches include the BFD method
\citep{bernstein14}, which derives a prior on the ensemble ellipticity distribution
using deeper fields,
and SNAPG \citep{herbonnet17}, a similar approach which builds ensemble shear estimates using
shear nulling. 

This paper is intended as a companion study to \citet{shearcat} (\citetalias{shearcat}),
where we present two shear catalogues derived from DES Y1 dataset.
It is also presented alongside a raft of other papers, which use both catalogues
and show them to be consistent in a number of different scientific contexts
\citep{shearcorr,gglpaper,y1massmaps,keypaper}
Containing 22 million and 35 million galaxies respectively,
these catalogues are the product of two independent maximum likelihood codes.
The first, called \imshape, implements simultaneous fits using multiple models
and we calibrate externally using simulations.
The second implements a Gaussian model fitting algorithm,
supplemented by shear response corrections using \metacal.
Whereas in \citetalias{shearcat} we focus on the catalogues themselves,
presenting a raft of calibration tests and a broad overview of the value-added data products,
here we use the same resources to explore a narrower topic:
the impact of image plane neighbours on shear measurement.
Specifically we use the image simulations described in \citetalias{shearcat},
from which the Y1 \imshape\ calibration is derived,
to explore the mechanisms for neighbour bias, 
and then propagate the results to mock shear two-point data
to investigate the consequences for weak lensing cosmology.
The results presented in this paper will be somewhat dependent
on the choice of measurement algorithm, selection cuts and the
configuration of the object detection code.
Unlike previous studies on this subject,
however, we make use of a highly realistic simulation and
measurement pipeline.
Our choices on each of aspects are realistic, if not unique, 
for a leading-edge cosmology
analysis.

It is worth remarking, however, that 
the tests described in this paper make use
of \imshape\ only, and should not be assumed to apply generically
to its sister Y1 \metacal~catalogue. 
A complementary set of tests using \metacal\
are presented in \S4.5 of \citetalias{shearcat}.

This paper is structured as follows. 
In Section \ref{section:theory} we briefly review the formalism of lensing,
and the observables discussed in this work.
In Section \ref{section:toy_model} we present a series of numerical calculations
using a toy model to characterise neighbour bias.
Section \ref{section:hoopoe_overview} decribes the simulated DES Y1 datasets,
generated using our \hoopoe\ simulator.
We test the earlier predictions under more typical observing conditions in Section \ref{section:hoopoe_main},
and extend them into a quantitative set of results using the more extensive 
Y1 \hoopoe\ dataset.
Section \ref{section:cosmology} then presents a numerical analysis designed to test
the cosmological implications of neighbour bias of the nature and magnitude found in our simulations.
We conclude in Section \ref{section:conclusion}. 


\section{The Shear Measurement Problem}\label{section:theory}

The problem of shape measurement is far more intricate than it might first appear.
Any cosmological analysis based on cosmic shear is reliant on a series of technical choices,
which can have a non-trivial impact on measurement biases, 
precision and cosmological sensitivity.
Specifically we must choose 
(a) how to parameterize each galaxy's shape, and which measurement method to use to estimate it,
(b) what selection criteria are needed to obtain data of sufficiently high quality for cosmology and 
(c) how biased is the measurement and what correction is needed? 
These choices should be made on a case-by-case basis, 
since the optimal solutions are dependent on a number of 
survey-specific factors. We discuss each briefly in turn below.

\subsection{Shape Measurement with \imshape}

The shape measurements upon which the following analyses are based make use of 
the maximum likelihood model fitting code 
\imshape\footnote{https://bitbucket.org/joezuntz/im3shape-git}
\citep{im3shape}.
It is a well tested and understood algorithm, which has since been used in a range of lensing studies 
\citep{svcosmology, whittaker15, kacprzak16, clampitt16}.
It was also one of two codes used to produce shear catalogues in the Science Verification (SV) stage 
and Year 1 of the Dark Energy Survey. 
We refer the reader to \citet{jarvis16} (hereafter \citetalias{jarvis16}) 
and \citetalias{shearcat} for the most recent modifications to the code. 

We use the definition of the flux signal-to-noise ratio of \citetalias{shearcat}, \citetalias{jarvis16} and \citet{great3}:

\begin{equation} \label{equation:snr_definition}
S/N \equiv \frac{\left ( \sum\limits_{i=1}^{N_\mathrm{pix}} f^\mathrm{m}_i f^\mathrm{im}_i / \sigma^2_i \right ) }
{\left ( \sum \limits_{i=1}^{N_\mathrm{pix}} f^\mathrm{m}_i f^\mathrm{m}_i / \sigma^2_i \right )^\frac{1}{2}}.
\end{equation}

\noindent
The indices $i=(1,2...N_\mathrm{pix})$ run over all pixels in 
a stack of image cutouts at the location of a galaxy detection.
The model prediction and observed flux in pixel $i$ are denoted
$f_i^\mathrm{m}$ and $f_i^\mathrm{im}$ respectively and $\sigma_i$
is the RMS noise. 
This signal-to-noise measure is maximised when the differences between the model and the image pixel fluxes
are small. 
Note that if the best-fitting model $f^\mathrm{m}$ is identical for two different postage stamps,
$S/N$ will favour the image with the greater total flux. 

A useful size measure, referred to as \rgp~is defined as
the measured Full Width at Half Maximum (FWHM) of the galaxy after PSF convolution,
normalised to the PSF FWHM. 
Real galaxy images are are not perfectly symmetric 
(i.e. size is not independent of azimuthal angle about a galaxy's centroid), and
single-number size estimates are obtained by circularising (azimuthally averaging) 
the galaxy profile and computing the weighted quadrupole moments of the resulting image.
For each galaxy we take the mean measured size across exposures.

\subsection{Shear Measurement Bias}
There are many ways bias can enter an ensemble shear estimate based on a population of galaxies. 
Although the list is not exhaustive, 
a handful of mechanisms are particularly prevalent,
and have been extensively discussed in the literature.
\begin{itemize}
  \item{\textbf{Noise Bias:} On addition of pixel noise to an image, the best-fitting parameters of a galaxy model will not scale linearly
  with the noise variance.
  This is as an estimator bias as much as a measurement bias, and results in an asymmetric, skewed likelihood surface 
  \citep{hirata03,refregier12,kacprzak12,miller13}.
  Any code which uses the point statistics of the distribution (either mean or maximum likelihood)
  as a single-number estimates of the ellipticity results in a bias.
  This is true even in the idealised case where the galaxy we are fitting can be perfectly decribed by our analytic light profile. 
  The bias is sensitive to the noise levels and also the size and flux of the galaxy, 
  and thus is specific to the survey and galaxy sample in question. 
  For likelihood-based estimates one solution
  would be to impose a prior on the ellipticity distribution
  and propagate the full posterior.
  However, the results can become dependent on the accuracy of that prior,
  and such codes require cautious testing using simulations \citep{bernstein14,simon16}}
  \item{\textbf{Model Bias:} In reality galaxies are not analytic light profiles with clear symmetries. 
  For the purposes of model-fitting, however, 
  we are constrained to use models with a finite set of parameters. 
  A model which does not allow sufficient flexibility to capture the range of morphological features
  seen in the images will produce biased shape measurements 
  \citep{lewis09,voigt10, kacprzak13}.}
  \item{\textbf{Selection Bias:} Even if we were to devise an ideal shape measurement algorithm, 
  capable of perfectly reconstructing the histogram of ellipticities in a certain population of galaxies, 
  our attempts to estimate the cosmological shear could still be biased. If a measurement step 
  prefers rounder objects or those with a particular orientation,
  the result would be a net alignment that could be mistaken as having cosmological origin. 
  In practice selection bias commonly arises from imperfect correction of PSF asymmetries 
  (eg \citealt{kaiser00, bernstein02}), 
  and the fact that many detection algorithms fail less frequently on rounder galaxies (\citealt{hirata04}). 
  It is such effects that make post facto quality cuts on quantities such as signal-to-noise or size 
  (both of which correlate with ellipticity) particularly delicate.}
  \item{\textbf{Neighbour bias:} In practice, galaxies in photometric surveys like DES are not ideal isolated objects.
  Rather, they are extracted from a crowded image plane using imperfect deblending algorithms.
  The term ``neighbour bias" refers to any biases in the recovered shear arising from the interaction between galaxies
  in the image plane. This can include both the direct impact on the per-galaxy shapes (e.g. \citealt{hoekstra16})
  and changes in the selection function (e.g. \citealt{hartlap10}).
  Neighbour bias is the subject of relatively few previous studies, and is the focus of this paper.}
\end{itemize}


\section{Toy Model Predictions}\label{section:toy_model}

To develop a picture of how image plane neighbours affect
shear estimates with \imshape, we build a simplified toy model. Using 
\blockfont{galsim}\footnote{https://github.com/GalSim-developers/GalSim}
we generate a $48\times48$ pixel postage stamp containing a single exponential
disc profile convolved with a tiny spherically symmetric PSF
(though we confirm that our results are insensitive to the exact size of the PSF).
We can then apply a small shear along one coordinate axis prior to convolution and
use \imshape\ to fit the resulting image.
In the absence of noise or model bias the maximum of the likelihood of
the measured parameters coincides exactly with the input values.
The basic setup then has four adjustable parameters: the flux and size of the galaxy
plus two ellipticity components, 
denoted $f_c$, $r_c$, $g_1^{tr}$ and $g_2^{tr}$.
Unless otherwise stated we fix these to the median values
measured from the DES Y1 \imshape~catalogue.
We do not model miscentering error
between the true galaxy centroid and the stamp centre.

It is worth noting that neither this basic model nor the more complex simulations 
that follow attempt to model spatial correlations in shear.
Even at different redshifts, a real neighbour-central pair share some portion of their line of sight. 
These spatial correlations will amplify the impact of blending, and are worthy of future investigation.
This is, however, likely a second-order effect of neighbours, and we postpone such study to a future date. 

\subsection{Single-Galaxy Effects}\label{subsection:single_object_effects}

To explore the interaction in single neighbour-galaxy instances
we introduce a second galaxy into the postage stamp,
convolved with the same nominal PSF.
This adds four more model parameters: neighbour size $r_n$, flux $f_n$, radial
distance from the stamp centre $d_{gn}$ and azimuthal rotation angle relative 
to the x coordinate axis $\theta$.
At this stage the neighbour has zero ellipticity.

\begin{figure*}
\begin{center}
\includegraphics[width=0.45\columnwidth]{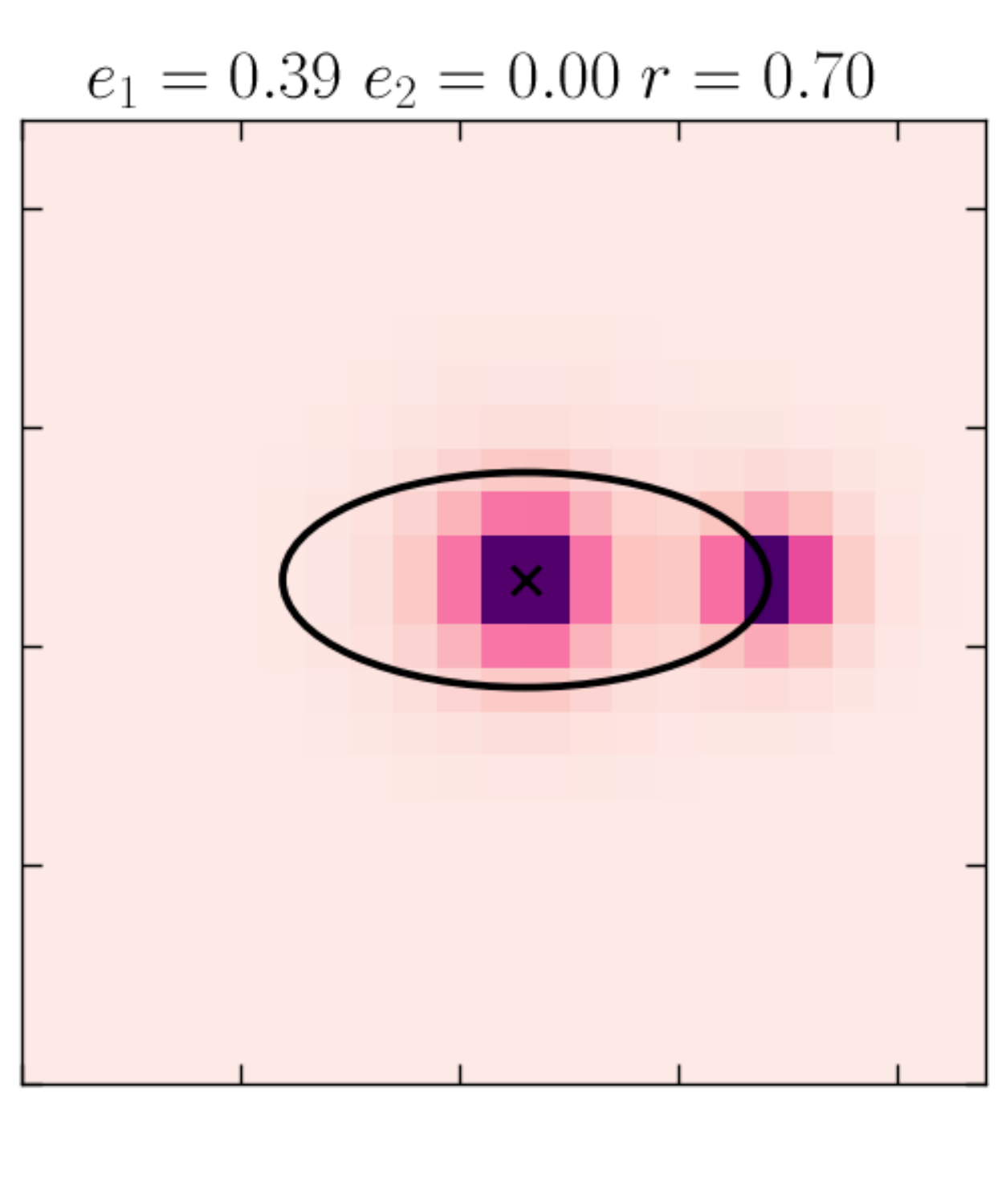}
\includegraphics[width=0.45\columnwidth]{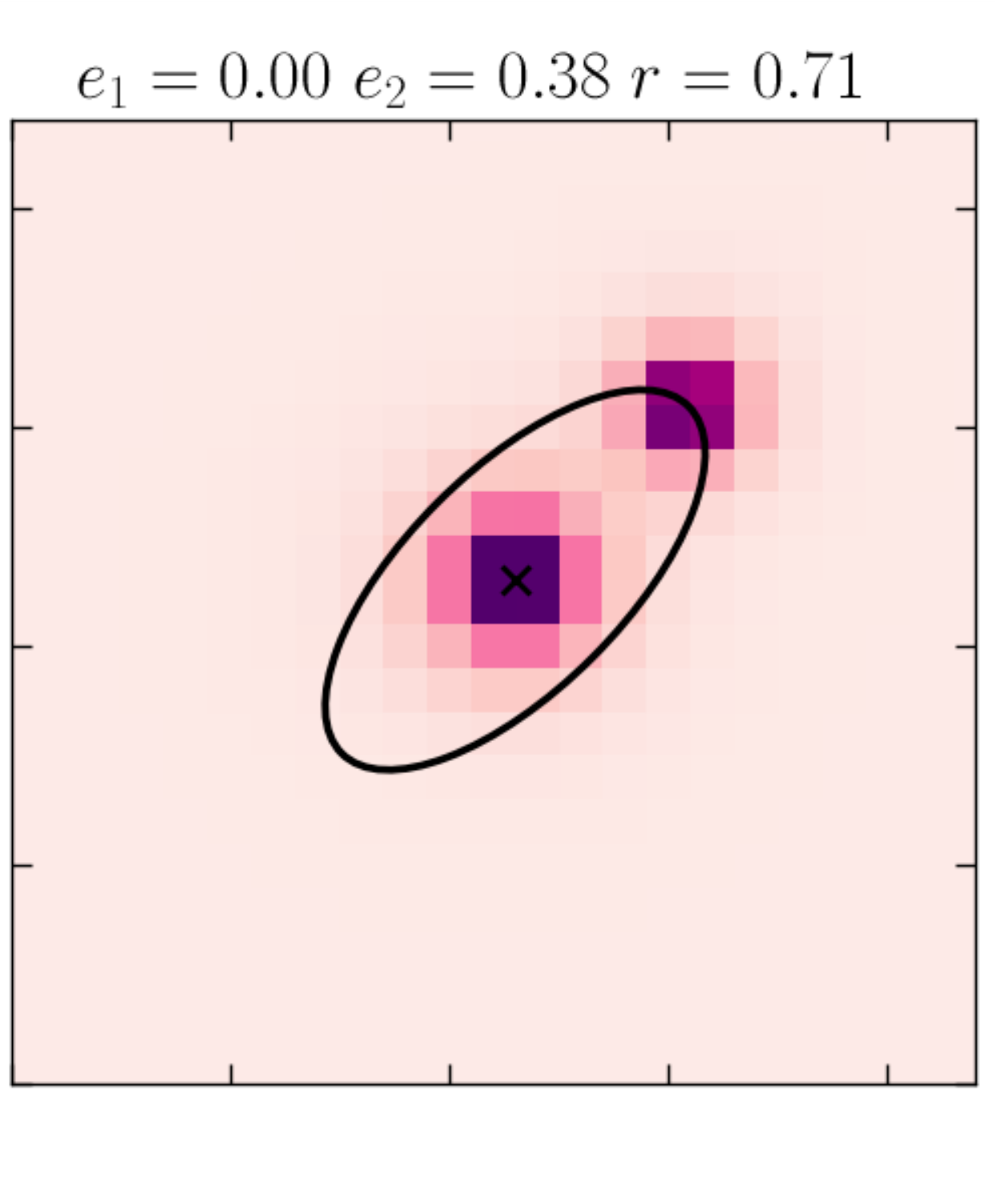}
\includegraphics[width=0.45\columnwidth]{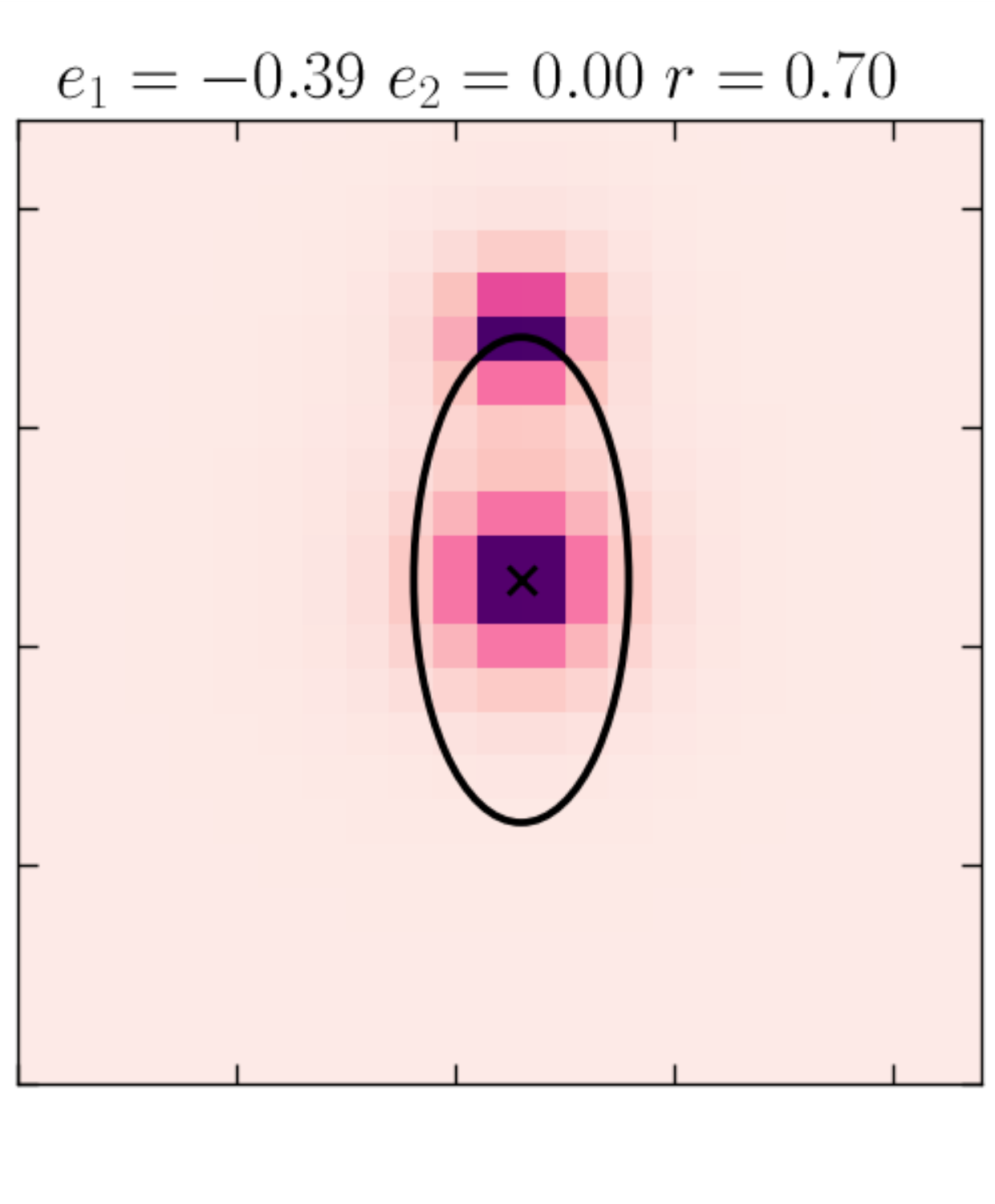}

\caption{Postage stamp snapshots of the basic two-object toy model described in Section \ref{section:toy_model}. 
The overlain ellipse shows the maximum likelihood fit to the image. 
The panels show three neighbour positions in the range $\theta = [0, \pi/2]$ rad. 
The best fit ellipticity and half light radius are shown above each image. In all cases the input values are 
$\mathbf{e}=(0,0)$, $r=0.5$ arcseconds.
}\label{fig:neighbour_positions}
\end{center}
\end{figure*}

\begin{figure}
\begin{center}
\includegraphics[width=0.85\columnwidth]{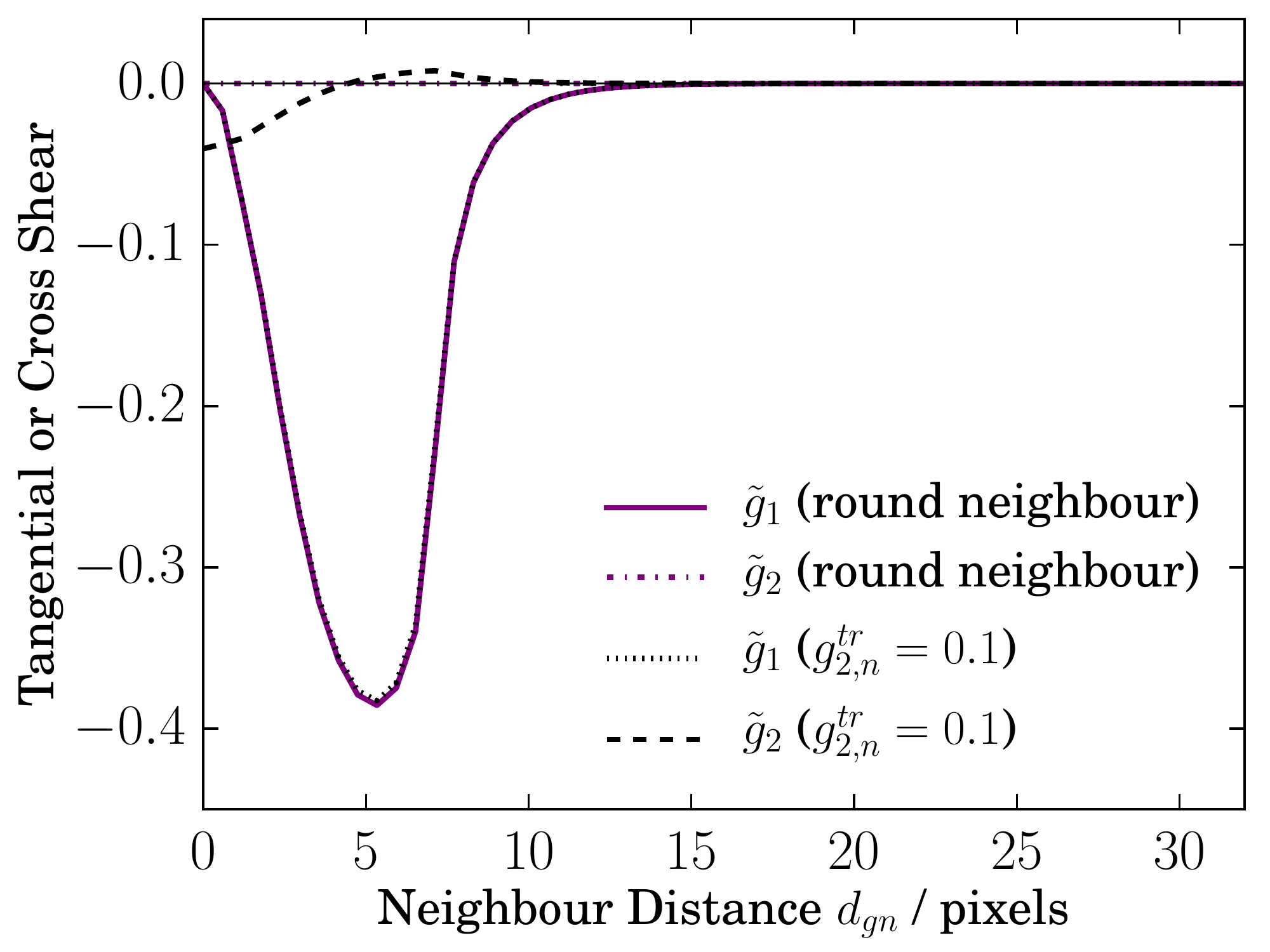}
\caption{Tangential shear measured using the numerical toy model described 
in Section \ref{subsection:single_object_effects} as a function of radial neighbour distance. 
The solid  purple line shows the shape component aligned 
with the central-neighbour separation vector and the dot-dashed line is measured
along axes rotated through $45^{\circ}$.
Note that the latter is smaller than $10^{-6}$ at all points on this scale.
The dashed and dotted black lines show the same ellipticity components when
the neighbour is sheared in the $e_2$ direction by $g_2=0.1$.
}\label{fig:etan-vs-dgn-toy_model}
\end{center}
\end{figure}

We show this setup at three neighbour positions in Fig. \ref{fig:neighbour_positions}.
Under zero shear, the system has perfect rotational symmetry, 
and the measured ellipticity magnitude $\tilde{g}(\theta|g_1^{tr}=0)$ is 
independent 
of $\theta$\footnote{Unless otherwise stated we fix the other model parameters to their fiducial values.}
As a first exercise, we generate a circular central galaxy with a circular Gaussian neighbour,
which is gradually shifted outwards from the stamp centre.
Following the usual convention for galaxy-galaxy lensing, tangential shear is defined such that
it is negative when the major axis of the measured shape is oriented radially towards the neighbour.
The measured two-component ellipticity shown by the solid and dot-dashed lines
in Fig. \ref{fig:etan-vs-dgn-toy_model}.
The decline in the measured tangential shear to zero at small separations is understandable,
as there is no reason to expect drawing one circular profile
directly atop another should induce spurious non-zero ellipticity.
In the regime of a few pixels, however, strong
blending can increase the flux of the best-fitting model. 

\begin{figure}
\begin{center}
\includegraphics[width=1.1\columnwidth]{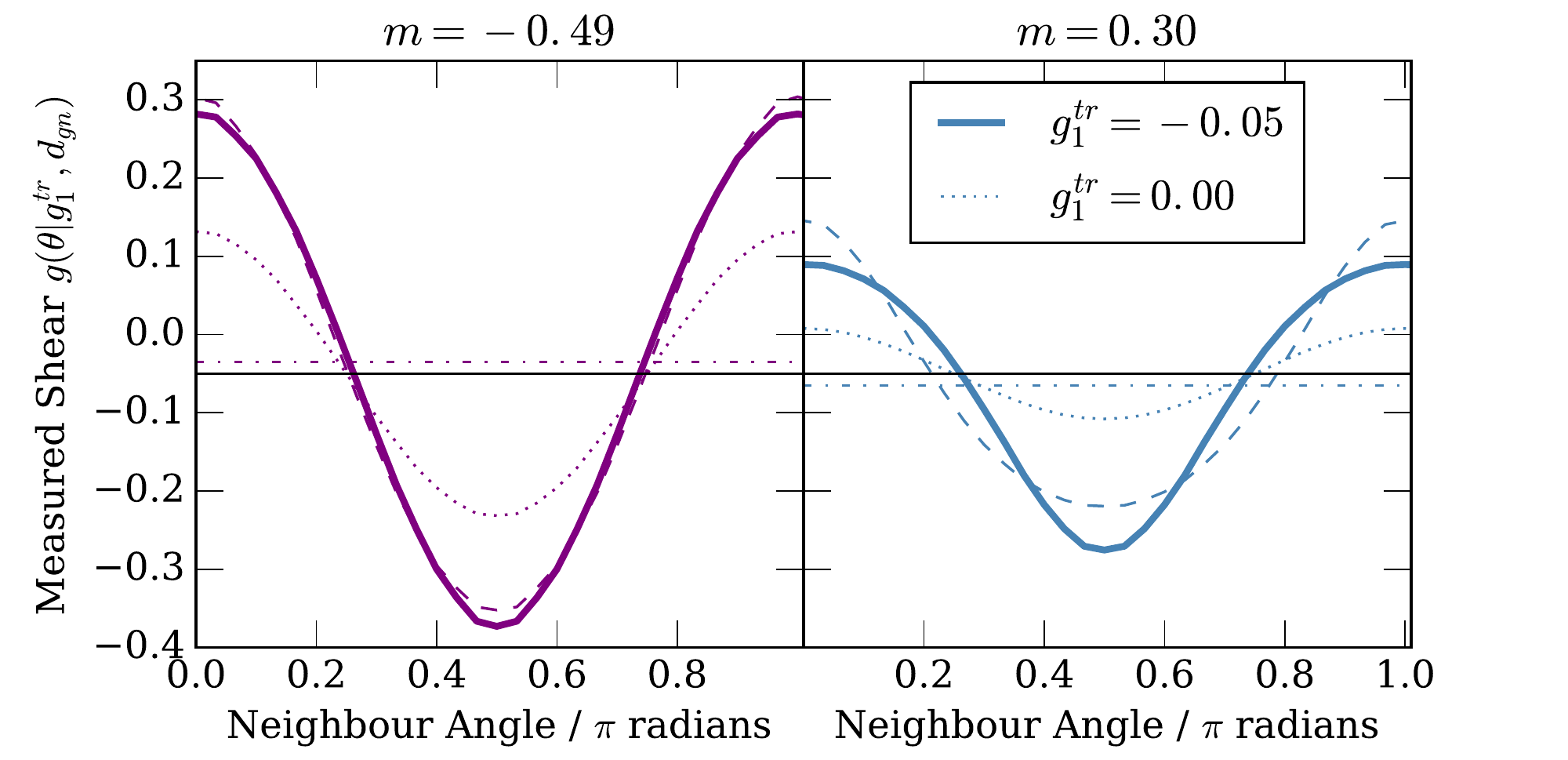}
\caption{Best-fit galaxy ellipticity as a function of neighbour position angle at 
fixed neighbour distance $d_{gn}$ from the toy model described in the text. 
The two panels (left, right) show the same central-neighbour system ($g^{tr}=-0.05$),
but with different $d_{gn}$ (7 and 8 pixels) and biases $m$ (shown atop each panel).
The solid line in each case is the recovered galaxy shape at each $\theta$,
and the integrated mean along this range is shown by the horizontal dot-dashed line.
The dotted lines show the zero-shear shape
(ie. the ellipticity that would be measured if the input shear were zero),
but shifted downwards such that the mean is at $-0.05$. 
Finally, to illustrate the (a)symmetry of the system we show the solid line flipped about 
$y=g^{tr}_1$ and shifted by $\pi/2$ radians as a dashed curve.
}\label{fig:g-vs-theta_neigh}
\end{center}
\end{figure}

Next, we repeat the calculation, now applying a moderate cross-component
shear to the neighbour ($g_2=0.1$).
The result is shown by the blue lines in Fig. \ref{fig:g-vs-theta_neigh}. 
Unsurprisingly the measured tangential shear is unaffected by a true
shear along an orthogonal axis. 
In cases where the objects share a large portion of their half-light radii,
we are fitting a strongly blended pair with a single profile, 
and the neighbour/central distinction becomes difficult to define.
The best-fitting ellipticity recovered from the blended image is not a pure measurement of either 
galaxy's shape; rather it is a linear combination of the two.
We repeat the zero-offset measurement using a range of neighbour fluxes and find that the best-fitting $e_i$
follows roughly as a flux-weighted sum over the two galaxies 
$\tilde{g}_i \approx (f_c g_{i,c}^{tr} + f_n g_{i,n}^{tr})/(f_n+f_c) $.

\subsection{Ensemble Biases}

\begin{figure}
\begin{center}
\includegraphics[width=0.6\columnwidth]{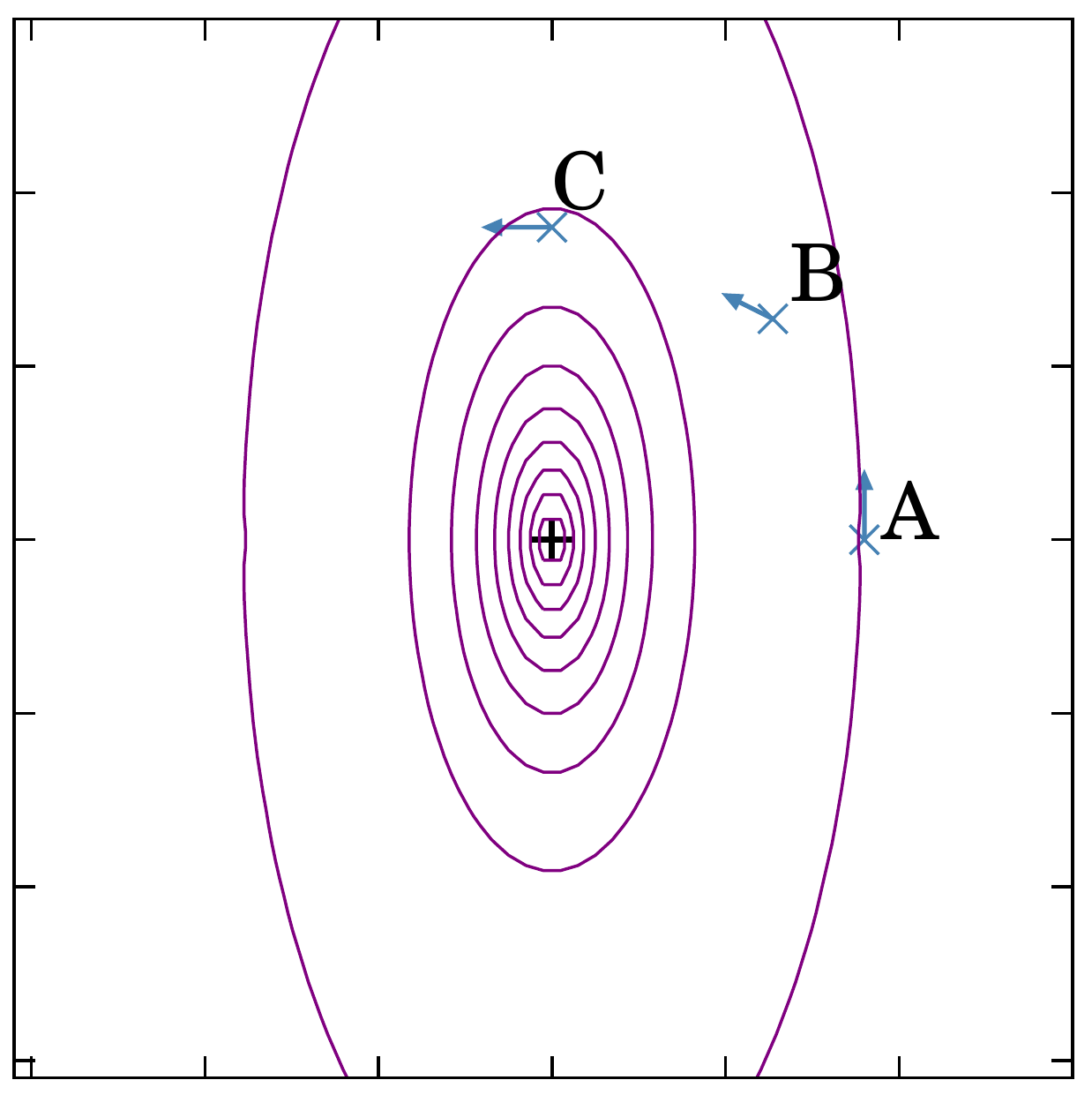}
\caption{Cartoon diagram of a neighbour-central system. 
The purple contours show the lines of constant flux in a \sersic~ disc profile with extreme negative ellipticity 
($g_1=-0.3$).
The blue crosses labelled A, B and C are points on a ring of equal distance from the centre of the profile.
The blue arrows show the local unit vector along a tangent to the ring.
}\label{fig:neigh_asymmetry}
\end{center}
\end{figure}

While useful for understanding what follows, 
the impact of neighbours on individual galaxy instances
is not particularly informative about the impact on cosmic shear measurements.
Even significant bias in the per-object shapes could average away over many galaxies with no residual impact on the recovered shear.
More important is the \emph{collective} response to neighbours.
To explore this we build on the toy model concept.
To estimate the ensemble effect, we measure a neighbour-central image at 70 positions on a ring of neighbour angles.
Again, under zero shear $g^{tr}=0$ the measured shape is constant in magnitude, and simply oscillates about 0 
with peaks of amplitude $|\tilde{g}(\theta|0, d_{gn})|$. 
This sinusoidal variation is shown by the dotted lines in Fig. \ref{fig:g-vs-theta_neigh}b at two values of $d_{gn}$
(7 and 8 pixels).
By averaging over a (large) number of neighbours one is effectively marginalising over $\theta$, which results in an 
unbiased measurement of the shear 
$\langle \tilde{g}(\theta|g^{tr}=0,d_{gn}) \rangle_{\theta} = g^{tr} = 0$.
A non-zero shear $g^{tr}\neq 0$, however breaks the symmetry of the system.
A galaxy sheared along one axis will not respond to a neighbour in the same way irrespective of $\theta$,
which can result in a net bias.
To show this we fix $g^{tr}=-0.05$ and proceed as before.
The solid lines in Fig. \ref{fig:g-vs-theta_neigh}b show the periodicity in the measured shear at two $d_{gn}$.
The mean value averaged over $\theta$ is shifted incrementally away from the input shear,
shown by the horizontal dot-dashed line. 
Specifically we should note that the peaks below $g^{tr}$ at $\pi/2$ and $3\pi/2$ radians are deeper and 
narrower than those above it.
The cartoon in Fig. \ref{fig:neigh_asymmetry} shows how this arises.
The purple lines are iso-light contours in a strongly sheared \sersic~disc profile ($g_1=-0.3$). 
Clearly rotating the neighbour from position A to C carries it from the 
relatively flat low wings of the central galaxy's light profile closer to the core. 
Perturbing an object about C by a small angle results in a much greater change
in the local gradient, 
$\bigtriangledown f_c (x,y)$
than doing the same about A. 
All other parameters fixed, an incremental shift along the blue tangent vector will have a larger impact at
$\theta=0$ than at $\pi/2$,
resulting in asymmetry in the width of
the positive and negative peaks in Fig. \ref{fig:neigh_asymmetry}.
The depth of the peak can be explained qualititatively by similar arguments.
At C a neighbour of given flux is closer to the centre of the light distribution and thus has a greater flux
overlap with the central galaxy than at A.
Naturally, then, one might expect neighbour A to have less impact than C.
Returning to Fig. \ref{fig:g-vs-theta_neigh}, we can see that the two effects are in competition.
Depending on the exact neighbour configuration, the simultaneous narrowing and deepening the negative peaks
can result in a bias in the neighbour-averaged ellipticity towards large or small values.

\begin{figure}
\begin{center}
\includegraphics[width=\columnwidth]{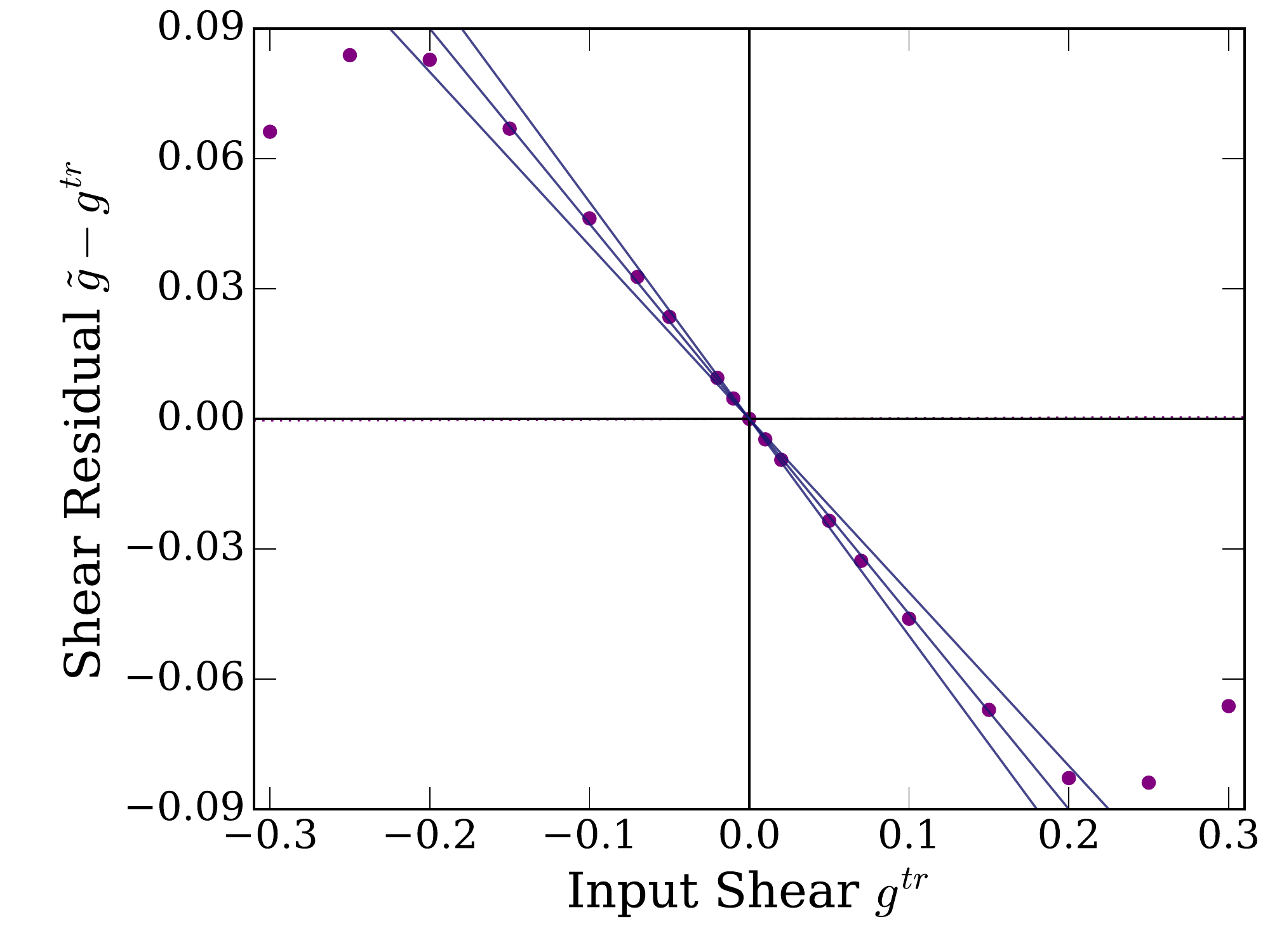}
\caption{Measured shear minus input shear plotted as a function of input shear. 
The purple points show the recovered $\tilde{g}_1$ from averaging over ring of 70 neighbour positions. 
The dark blue lines show the linear relation $\tilde{g}-g^{tr} = mg^{tr}$ at $m=(-0.4, -0.45, -0.5)$.
The dotted line shows what would be measured using the same central profile
in the absence of the neighbour, and is near indistinguishable from the x axis line on all points
within this range of $g^{tr}$.
}\label{fig:dg-vs-g}
\end{center}
\end{figure}

The level of this effect will clearly correlate with the magnitude of the shear,
and so induce a multiplicative bias. 
To illustrate this point the above exercise is repeated with a range of different input shears.
The results for our fiducial setup are shown in in Fig. \ref{fig:dg-vs-g}.
Each point on these axes corresponds to a ring of neighbour positions for a given input shear.
The equivalent measurements without the neighbour are indistinguishable from the x axis.  
At small shears, the neighbour induced bias $\tilde{g}-g^{tr}$ is well aproximated as a linear in $g^{tr}$.
We leave exploration of the possible nonlinear response at large ellipticities for future investigations.
Though the above numerical exercise demonstrates that it is \emph{possible} for significant multiplicative bias
to arise as a result of neighbours, it does not make a clear prediction of the magnitude or even the sign.
Indeed, our toy model is effectively marginalised over $\theta$,
but there is nothing to guarantee that fixing the other neighbour parameters to the median measured values
is representative of the real level of neighbour bias in a survey like DES.
Motivated by this observation we add a final layer of complexity to the model, as follows.
A single neighbour-central realisation is created as before, defined by a unique set of model parameters.
Now, however, the values of those parameters $\textbf{p}=(d_{gn}, f_n, r_n, f_c, r_c)$ are drawn randomly from the DES data. 
As these quantities will, in reality, be correlated
we sample from the 5-dimensional joint distribution rather than each 1D histogram individually. 
We then fit the model at 70 neighbour angles and two input shears 
$g_\pm = \pm 0.05$ (a total of 140 measurements),
and estimate the multiplicative bias as a two-point finite-difference derivative:

\begin{equation}\label{eq:mc_tm_m}
m + 1= \frac{ \langle \tilde{g}(\theta|g_+) \rangle_{\theta} - \langle \tilde{g}(\theta| g_-) \rangle_{\theta}} {g_+ - g_-}.
\end{equation}

\begin{figure}
\begin{center}
\includegraphics[width=\columnwidth]{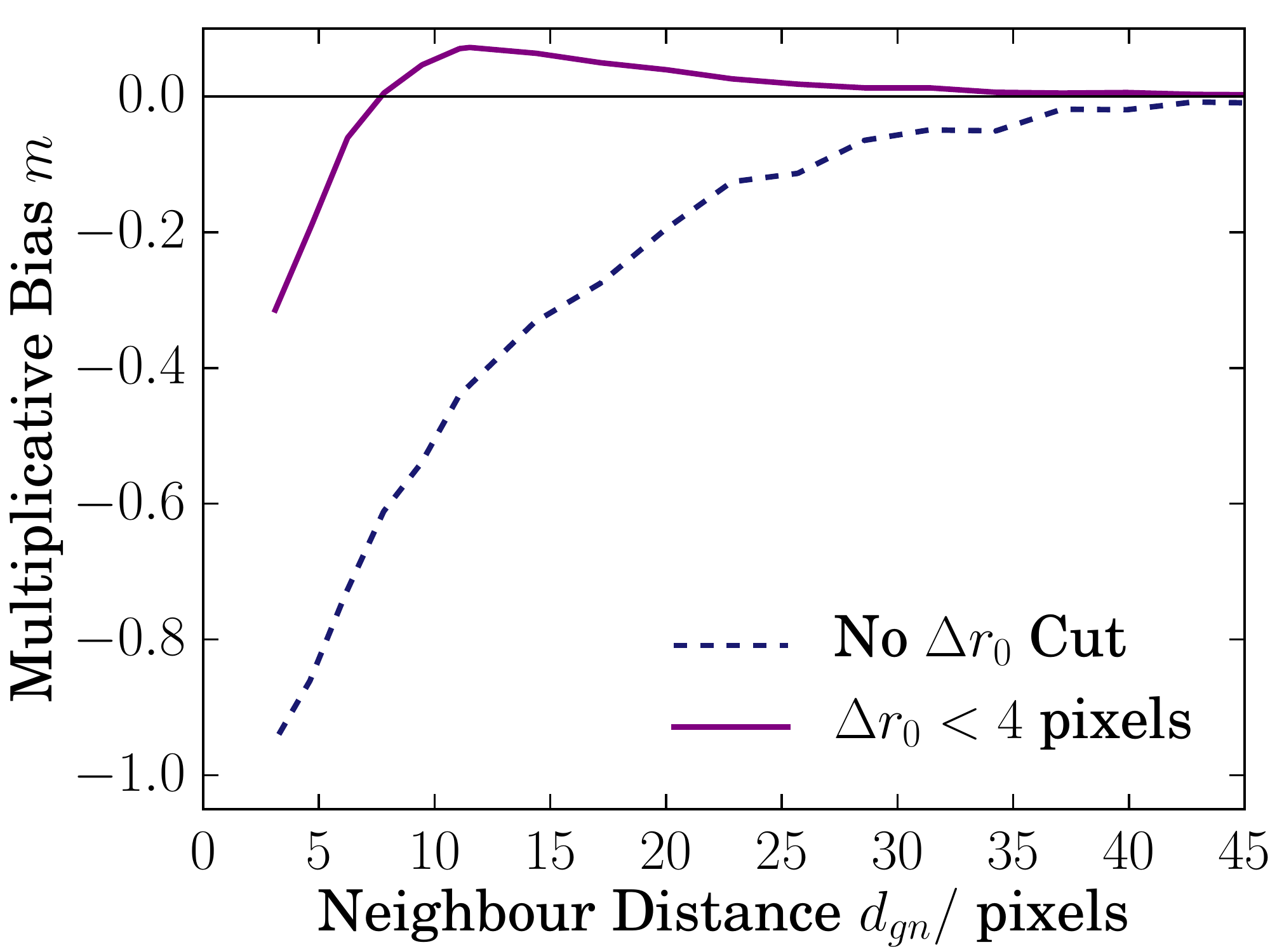}
\caption{Multiplicative bias estimated using the Monte Carlo toy model described in the text.
For each neighbour realisation, defined by a particular distance, flux and size we compute 
the average of the measured ellipticity components over 70 rotations on a ring of neighbour angles. 
To estimate the bias we perform this averaging twice at two non-zero shears, $g_+$ and $g_-$, and 
compute the finite-difference deriviative using equation \ref{eq:mc_tm_m}.
The dashed thin blue line shows the result of using all measurements, while the bold purple line
has a cut based on the offset between the centroid position of the best-fitting model and the stamp
centre.
}\label{fig:dgn-vs-m-mc_toymodel}
\end{center}
\end{figure}

\noindent
This process is repeated to create 1.33M unique toy model realisations.
Binning by neighbour distance we can then make a rough prediction for the level of neighbour-induced bias
and the angular scales over which it should act. 
The result is shown in Fig. \ref{fig:dgn-vs-m-mc_toymodel},
where full results using all model realisations are indicated by the dashed blue line.
The majority of cases yield a negative bias, particularly at low neighbour separation
(referring back to Fig. \ref{fig:neigh_asymmetry}, the broadening of the peak around position A 
dominates over the increased flux overlap at C).
In the real data, of course, we apply a quality based selection and \"uberseg object masking
\citepalias{jarvis16},
both of which are neglected here. 
We can, however, test the impact of selecting on fitted quantities that respond to neighbour bias. 
Imposing a flat prior on the centroid offset $\Delta r_0 = ( x_0^2+  y_0^2)^{\frac{1}{2}}$ 
(i.e. discarding randomly generated model realisations where the galaxy centroid is 
displaced from the stamp centre by more than a fixed number of pixels)
changes the shape of this curve significantly, as illustrated by the thick purple line. 

We can understand the difference between the results with and without the centroid cut as a form of
selection bias, whereby the cut preferentially removes toy model realisations in which the neighbour 
is bright relative to the central galaxy.
At any given $d_{gn}$ we are left with a relative overrepresentation of galaxies with $f_n/f_c \ll 1$.
Faint neighbours, which in reality tend to be compact high redshift objects, have little impact when they sit
on the outskirts of the central profile 
(A in the cartoon picture in Fig. \ref{fig:neigh_asymmetry}; the regime which produces negative $m$). 
The same faint galaxy has a stronger impact if it is rotated to a position closer to the centre of the 
central's flux profile.
Thus one might expect a selection on $\Delta r_0$ to make the mean $m$ in a particular bin less negative 
(or even positive) by preferentially removing brighter galaxies.

\section{\hoopoe~Image Simulations}\label{section:hoopoe_overview}

\begin{figure*}
 \centering
\includegraphics[width=1.25\columnwidth]{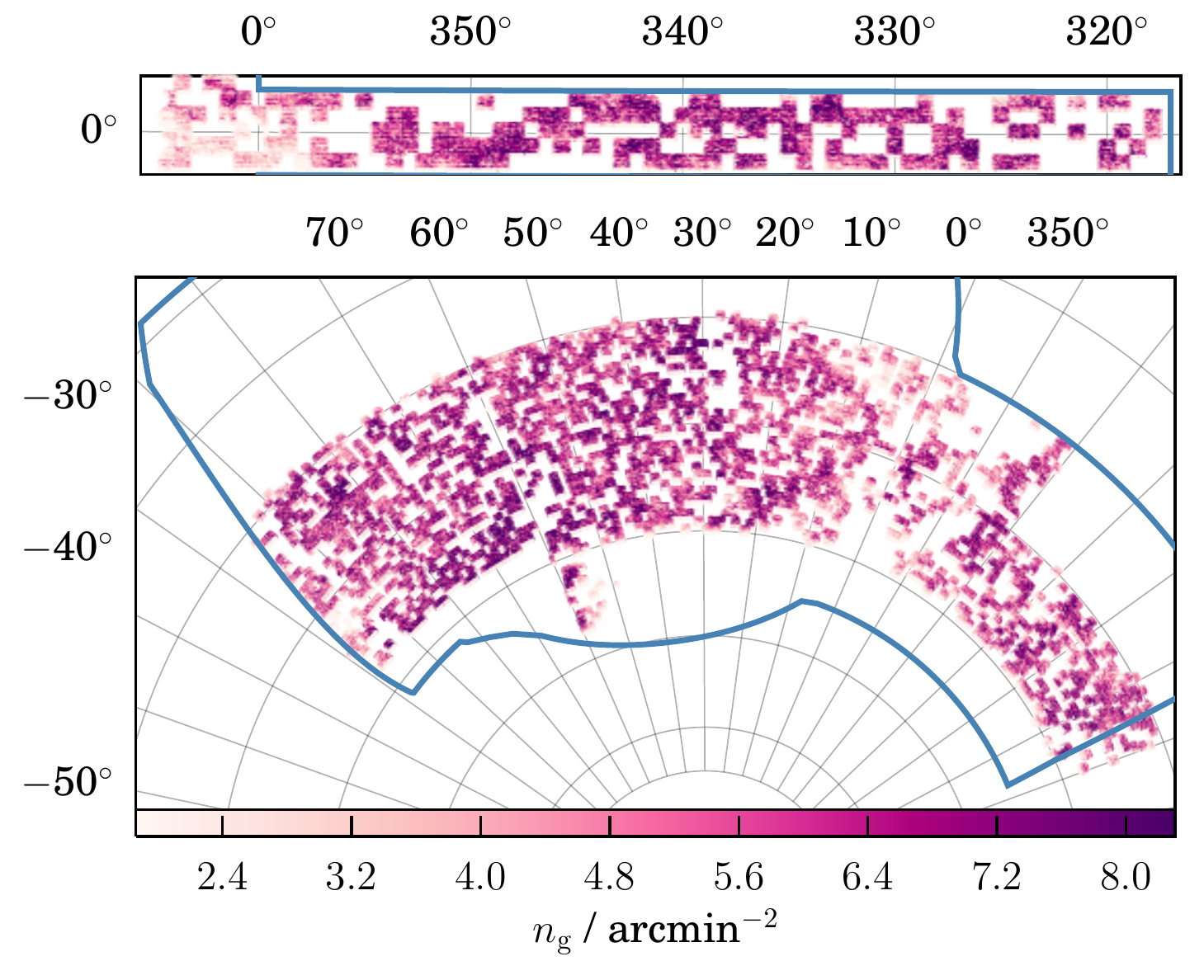}

(a)
\includegraphics[width=0.75\columnwidth,height=0.75\columnwidth ]{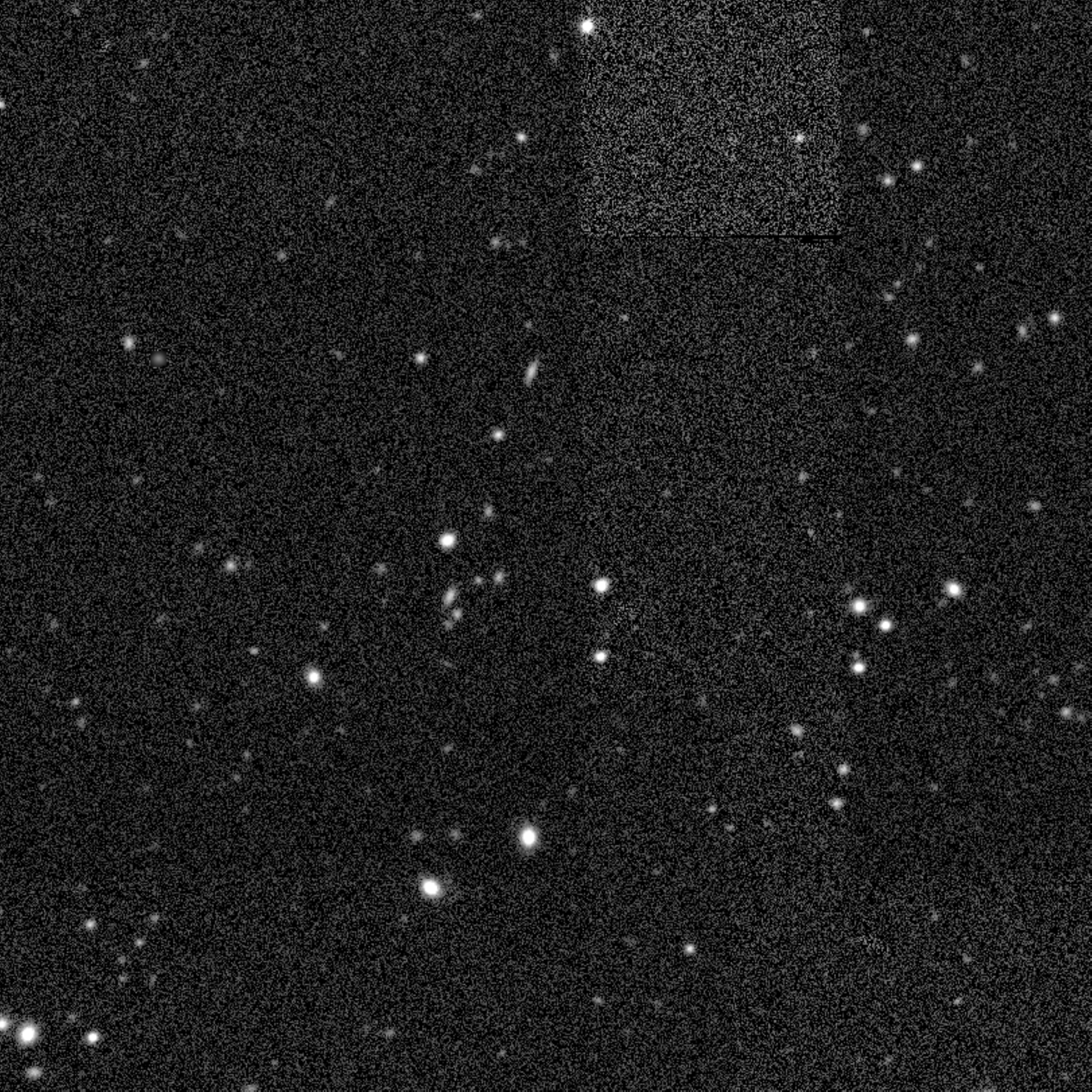}
(b)
\includegraphics[width=0.75\columnwidth,height=0.75\columnwidth]{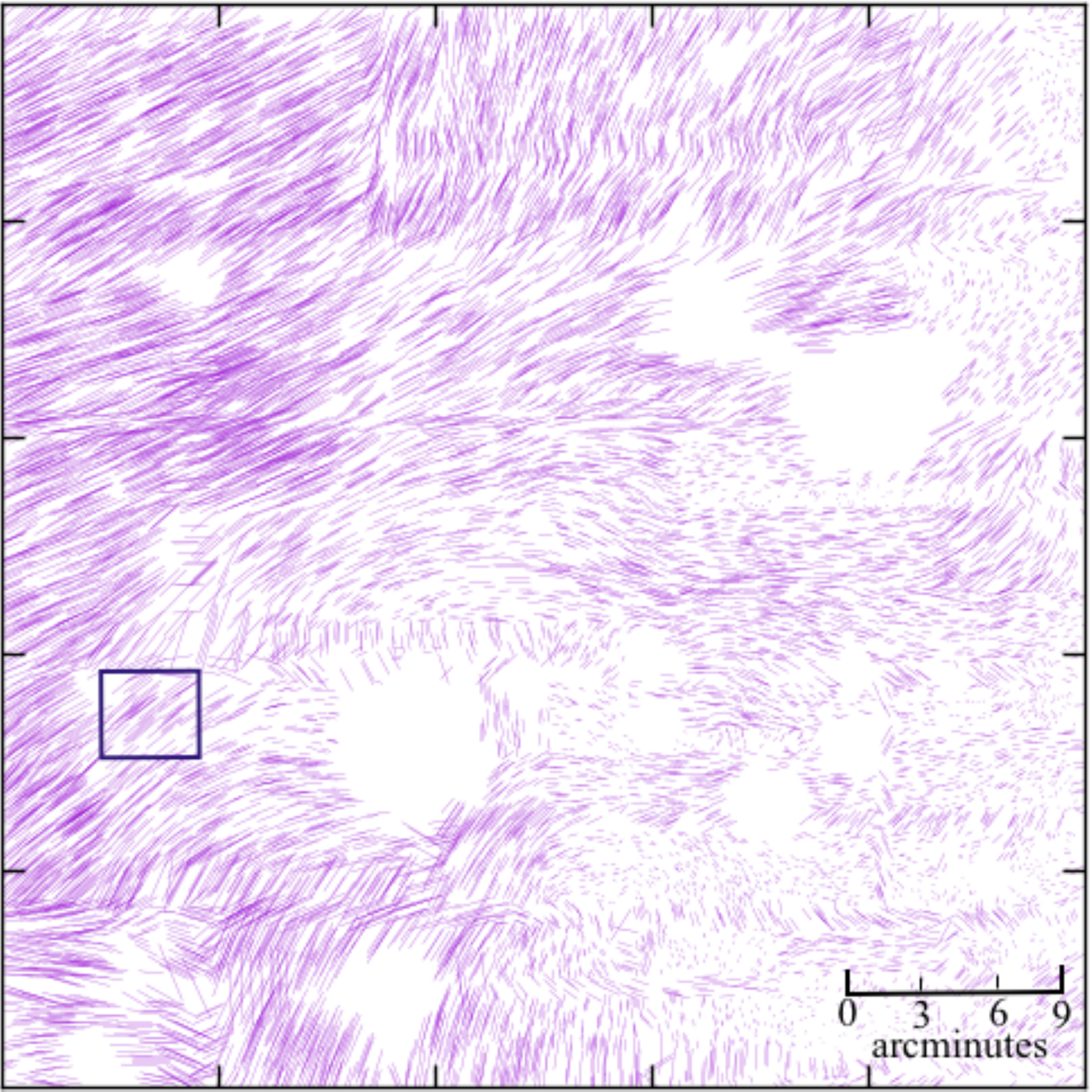}
\caption{\textbf{Top:} The projected footprint of the simulated survey,
visualised using the \blockfont{skymapper} package$^{a}$.
The colour indicates the local raw number density in \blockfont{healpix} 
cells of $\mathrm{nside} = 1024$.
The axes shown are right ascension and declination in units of degrees. 
The full simulation comprises 1824 $0.73 \times 0.73$ degree tiles drawn randomly from the DES Y1 area.
The solid blue line indicates the bounds of the planned area to be covered by the complete Y5 dataset. 
\textbf{Bottom:} A random tile (DES0246-4123) selected from the \hoopoe~area. 
The left panel (a) shows a square subregion of approximately $4\times4$ arcminutes. 
The right hand panel (b) shows a PSF whisker plot covering the full $0.73\times0.73$ tile.
The length and orientation of each line represents the magnitude and position angle of the spin-2 PSF 
ellipticity at that position. 
Only galaxies which pass \imshape~quality cuts are shown. 
The white patches show the spatial masking inherited from the \blockfont{gold} catalogue,
and correspond to the positions of bright stars in the parent data.
$^a$https://github.com/pmelchior/skymapper
}\label{fig:sky_map_nexp}
\end{figure*}

In this section we provide a brief overview of the simulation pipeline. 
The process is the same as that described in \S4 of \citetalias{shearcat},
and we refer the reader to that work for more detail.
The end point of the pipeline is a cloned set of survey images 
with many of the observable characteristics of a chosen set of parent images,
but for which we know the input noise properties and galaxy population pefectly.
The simulated images inherit the pixel masking, PSF variation and noise maps 
measured from the progenitor data. 
Each simulated galaxy is then inserted into a subset of overlapping exposures
and into the coadd at the position of a real detection in the DES Y1 data.
Object detection is rerun on the new coadd images 
and galaxy cutouts and new segmentation masks are extracted and stored in the MEDS
format decribed by \citetalias{jarvis16}.
The mock survey footprint is shown in Fig. \ref{fig:sky_map_nexp}.
In the lower panels we show an example of a simulated coadd (left)
and the spatial variation in PSF orientation within the same image (right). 

\subsection{Parent Data}

We use reduced images from Year One of the Dark Energy Survey 
(DES Y1; \citealt{diehl14}) as input to the simulations discussed in this paper.
The Dark Energy Survey is undertaking a five year programme with 
the ultimate aim of observing $\sim5000$ square degrees of the southern sky to 
$\sim 24$th magnitude in five optical bands, \emph{grizY},
covering $0.40-1.06$ microns.
The dataset is recorded using a 570 megapixel camera called DECam \citep{flaugher15},
which has a pixel size of $0.26$ arcseconds.
In full it will consist of $\sim10$ interwoven sets of exposures 
in the \emph{g}, \emph{r}, \emph{i}, \emph{z} and \emph{Y} bands.

The Y1 data were collected between August 2013 and February 2014,
and cover a substantially larger footprint than the preliminary
Science Verification (SV) stage
at 1500 square degrees, albeit to a reduced depth.
Details of the reduction and processing are presented in \citetalias{shearcat}. 
Our \hoopoe\ simulations use a selection of the total 3000 $0.75\times0.75$ degree coadded patches
known as ``tiles".

\subsection{Input Galaxy Selection} \label{subsection:input_gals}

For populating the mock survey images a sample of real galaxy profiles
from the HST COSMOS field,
imaged at significantly lower noise 
and higher resolution
than DES by the Hubble Space Telescope
Advanced Camera for Surveys (HST ACS) \citep{scoville07}. 
The COSMOS catalogue extends significantly deeper than the Y1 detection limit of 
$M_\mathrm{r,lim} =  24.1$, extending to roughly $27.9$ mag in the SDSS $r$-band.
A main sample for our DES Y1 simulations is defined by imposing a cut at $<24.1$ mag.

Since the DES images do not cut off abruptly at 24th magnitude,
in reality they contain a tail of fainter galaxies that contribute flux
are not identifiable above the pixel noise.
To assess the impact of these objects on shape measurements in Y1,
we simulate a population of sub-detection galaxies
in addition to the main sample.
In brief we use the full histogram of COSMOS magnitudes to estimate the number of faint 
galaxies within a given tile. 
The required profiles are selected randomly from the faint end
of the COSMOS distribution.
Each undetected galaxy is paired with a detection,
and inserted at a random location within the overlapping bounds of the 
same (subset of) single-exposure images.
A more detailed description of this process can be found in \citetalias{shearcat}.

If these galaxies were present in the data they would enter the background flux calculation, 
and thus the subtraction applied would change due to their presence. 
Since the simulation pipeline produces images effectively in a post-background subtraction state this is not captured. 
To test this we rerun the \se~background calculation on a handful of tiles drawn with and without the faint galaxies.
The impact was found to be well approximated as a uniform shift in the background correction.
A flux correction equal to the pixel-averaged flux of the sub-detection galaxies over 
each image plane is, then, applied to postage stamps prior to shape measurement.

In reality
the overdensity of sub-threshold galaxies will be coupled to the density of detectable objects,
which is clearly not the case in our simulations.
To gauge the impact of this we perform the
following test.
Each tile is divided into a $6\times6$ grid, and the mean multiplicative
bias is calculated in each sub-patch.
We bin sub-patches according to the ratio $f_\mathrm{faint} \equiv N_\mathrm{faint}/N_{det}$, 
or the total number of faint galaxies relative to the number of detectable ones.
The impact is significant, but not leading order;
excluding patches outside the range $0.9< f_\mathrm{faint} < 1.1$
induces a shift of $\Delta m \sim -0.005$.

An independent noise realisation is generated for each exposure using the weight map from the parent data. 
We simulate the noise in each pixel by drawing from a Gaussian of corresponding width. 
The coaddition process is not rerun, but rather we compute an independent noise field by drawing the flux 
in each pixel from a zero-centred Gaussian of width determined by the measured variance in that pixel.

\subsection{Neighbour-Free Resimulations}\label{subsection:resims}

\begin{figure}
\includegraphics[width=\columnwidth]{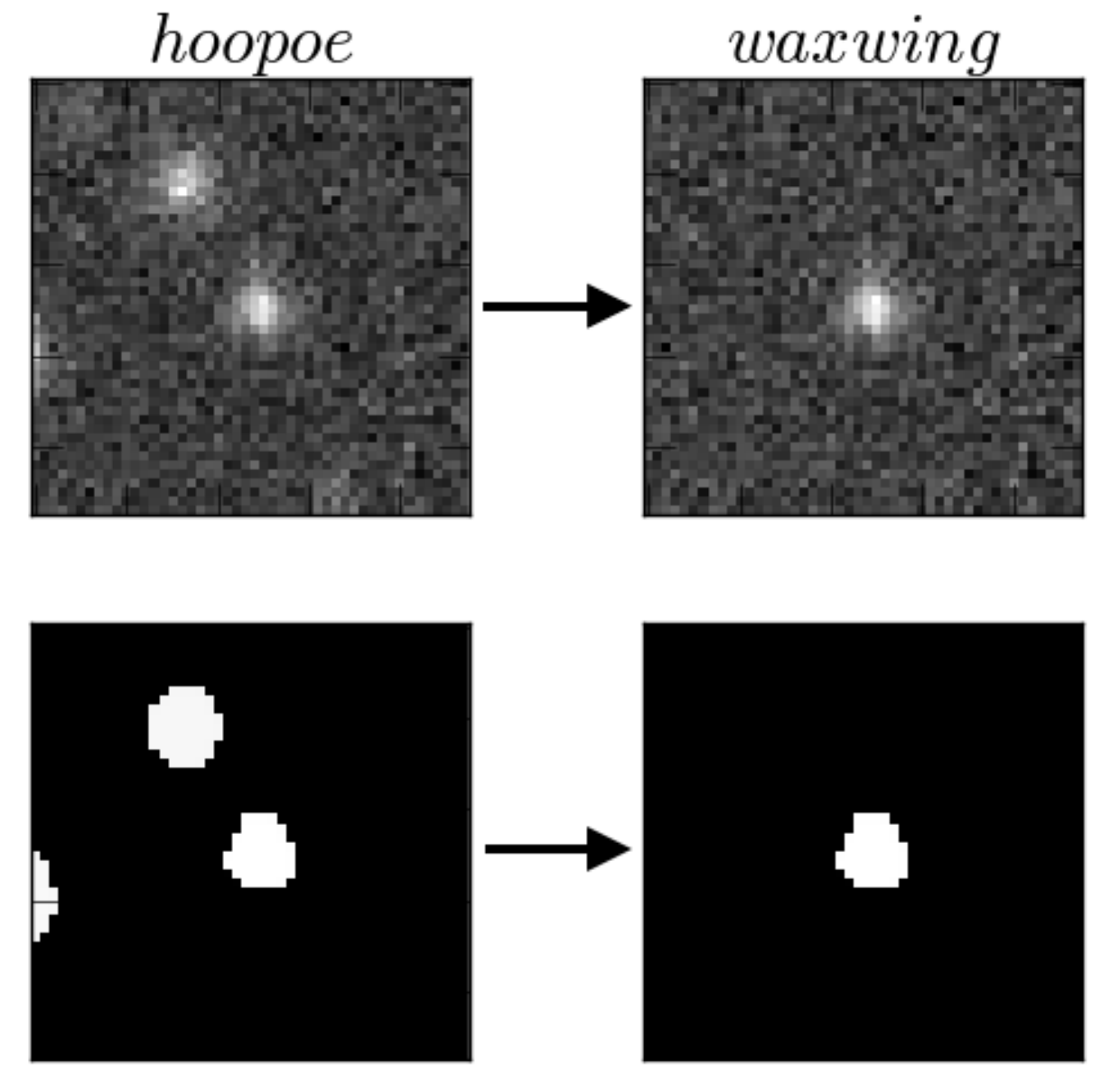}
\caption{An example of an object in the main DES Y1 calibration simulation and the neighbour-free 
resimulation. 
The upper panels show the coadd cutout in the original simulated images (left, labelled \hoopoe) and in 
the neighbour-subtracted version (right, labelled \waxwing).
The lower panels are the segmentation masks for the same galaxy. A number of neighbours, both masked 
(upper left and centre left) 
and unmasked (lower right) are visible within the stamp bounds.
}\label{fig:waxwing_example}
\end{figure}

For the purpose of untangling the impact of image plane neighbours we use the simulated  
\hoopoe~images to create a new spin-off dataset. 
In a subset of a little over 500 tiles we store the (convolved) input profile for each object and the noise-only cutout,
taken from the same position in the image plane prior to objects being drawn.
By adding together these two components we can generate a suite of spin-off MEDS files,
which are equivalent to the results of a simpler neighbour-free simulation 
(eg  \citealt{miller13}, \citetalias{jarvis16}).
The pixel noise realisation, COSMOS selection and input shears,
however, are identical to the progenitor \hoopoe~simulations. 

We will call this process ``resimulating", and the basic concept is illustrated in Fig. \ref{fig:waxwing_example}. 
The 506-tile set of neighbour-free data are named the \waxwing\ resimulations. 
Finally the (now empty) segmentation masks corresponding to the subtracted neighbours are also removed. 
In subsequent \imshape\ runs on these data we ignore the \se~flags obtained from the main simulations.


\section{Quantifying Neighbour Bias with \hoopoe}\label{section:hoopoe_main}

Equipped with qualitative predictions from Section \ref{section:toy_model},
we now turn to the question of neighbour bias in the more complete 
simulations described in Section \ref{section:hoopoe_overview}.
The mock survey was designed to capture as much of the complexity of shape measurements 
on real photometric data as possible.
We refer to Section \ref{section:hoopoe_overview} of this paper for a short overview and to 
\S 5 of \citetalias{shearcat} for a more detailed discussion of the simulation pipeline 
and validation tests. 
The simulated galaxy catalogue used in the following is identical to the one used to 
calibrate the DES Y1 \imshape~catalogue, including quality cuts and selection masks.

\subsection{Single-Galaxy Effects}

The most straightforward way to assess the impact of neighbours on individual shape measurements
in our simulations is to rotate the measured shapes into a frame defined by the central-neighbour separation vector.
Whereas in the earlier toy model we had only one neighbour per galaxy,
we now have a crowded image plane containing many objects simultaneously.
For simplicity, in the earlier case we included no masking.
For \hoopoe~we wish to mimic the process of shape measurement on
real data as closely as possible. We generate new segmentation maps by
running \se~on the simulated images, 
and incorporate them into our shape measurements using
the {\"u}berseg algorithm (\citetalias{jarvis16}).  
Each simulated galaxy is allocated a nearest neighbour using a \emph{k}-d tree
matching algorithm
constructed on the coadd pixel grid using every galaxy simulated at $r$-band magnitude $M_\mathrm{r}<24.1$.
The quantities $d_{gn}$ and $\theta$ are now redefined slightly as nearest-neighbour distance and angle. 
We define the tangential shear of a galaxy relative to its nearest neighbour using the standard convention,

\begin{equation}
e_{+} = - \left [ e_1 \cos (\theta) + e_2 \sin (\theta) \right ],
\end{equation}

\noindent
and the cross shear

\begin{equation}
e_{\times} = - \left [ e_2 \cos (\theta) - e_1 \sin (\theta) \right ].
\end{equation}

\begin{figure}
\includegraphics[width=1\columnwidth]{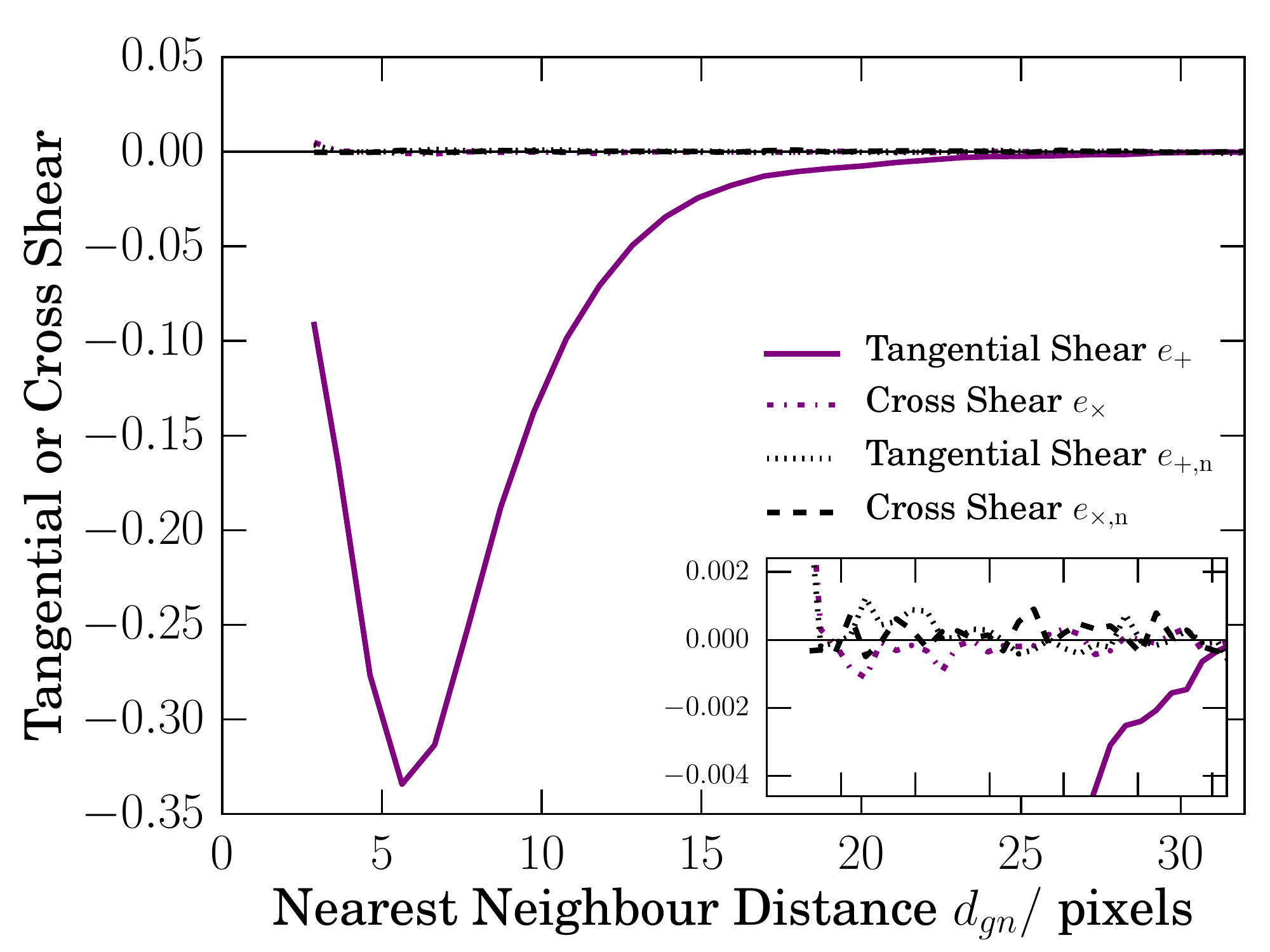}
\caption{Tangential shear around image plane neighbours in the full \hoopoe~simulation. 
The purple solid line shows the mean component of the measured galaxy shapes radial to 
the nearest image plane neighbour. 
Dashed blue shows the component rotated by $45^{\circ}$, which we have no reason to expect 
should be non-zero.
The dotted and dot-dash lines show the measured ellipticity components when rotated into a 
coordinate frame defined by the major axis of the neighbour. 
The inset shows the same range in $d_{gn}$ (the x-axis tick markers are the same),
but with a magnified vertical axis.
}\label{fig:tangential-shear-around-neighbours}
\end{figure}

\noindent
Note that negative values of $e_{+}$ imply a net tangential alignment of the measured shapes towards neighbours. 
By analogy, we define $e_{1,\mathrm{n}}$ and $e_{2,\mathrm{n}}$, which are the measured ellipticity components,
rotated into a reference frame defined by the major axis of the neighbour.
Non-zero $e_{i,\mathrm{n}}$ would indicate leakage of the neighbour's shape into the measurement,
which might conceivably be induced by inadequate deblending of very close neighbours or by extensive non-circular masking.
We first divide the main simulated catalogue into bins according to $d_{gn}$, and measure the tangential shear 
about nearest neighbours in each bin. 
The result is shown by the purple curve in Fig. \ref{fig:tangential-shear-around-neighbours}.
Note that the statistical uncertainty is within the width of the line in all bins. 
The results here show qualitative agreement with the numerical predictions in 
Fig. \ref{fig:g-vs-theta_neigh}.
As we found earlier, the exact shape of this curve is sensitive to the properties
of both the neighbour and the central galaxy.
Despite small differences, the range of variation is comfortably within the scale of the
postage stamp for the bulk of galaxies in DES Y1.
Repeating the measurement, rotated into the plane of the neighbour shape results in
the dotted and dot-dash lines in this figure.
As noted above, there are not necessarily reliable ellipticity measurements for each
neighbour, so we instead use the sheared input ellipticities.
Both components of $e_{i,\mathrm{n}}$ are seen to be negligible over all scales.

\subsection{Neighbour Ensemble Biases}

\begin{figure}
\includegraphics[width=1\columnwidth]{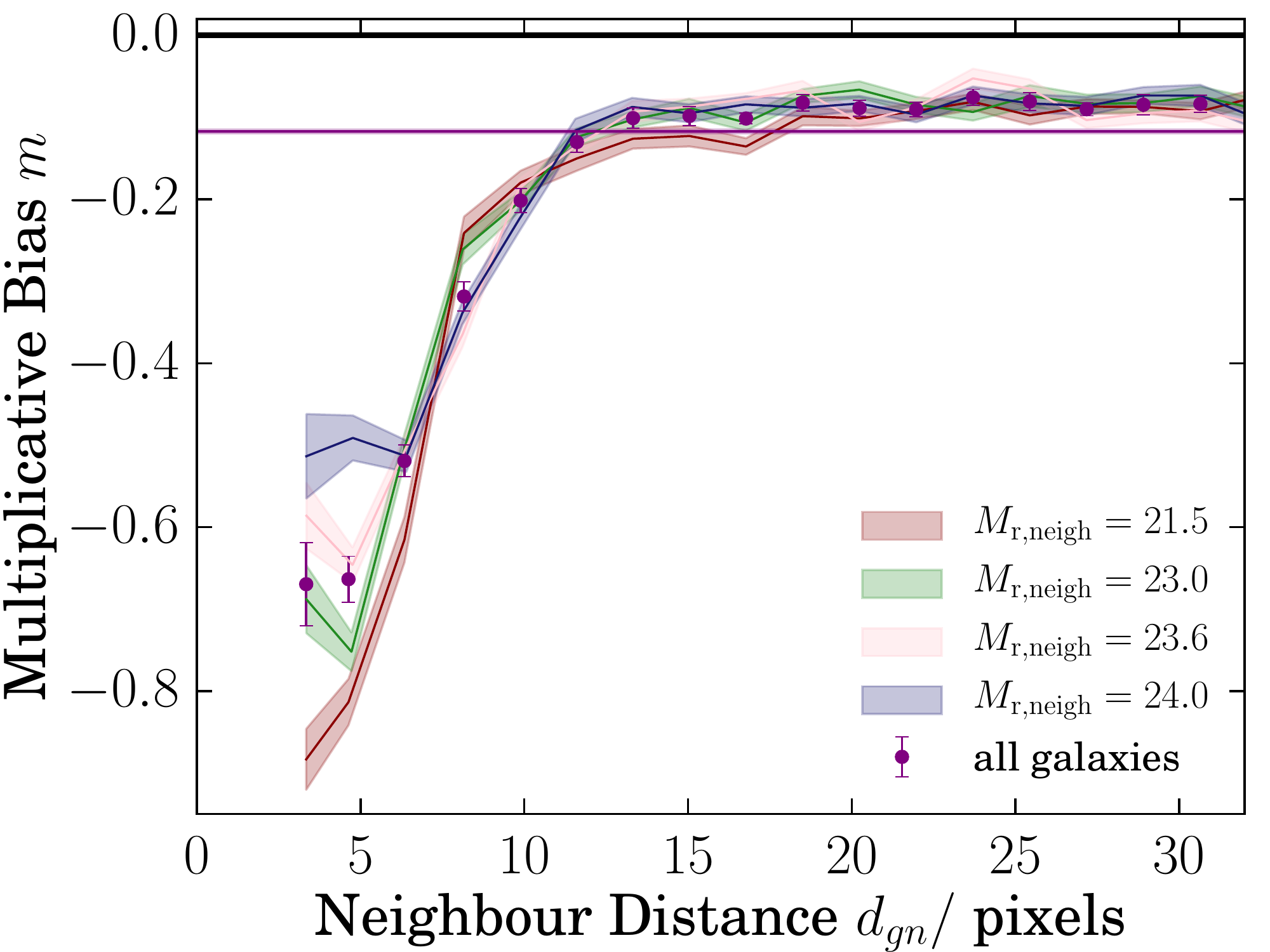}
\caption{Multiplicative bias as a function of separation from the nearest image plane neighbour.
The purple points show the bias calculated in bins of neighbour distance using the main \hoopoe\ simulated shape catalogue.
The coloured bands show the same dataset divided into four equal-number bins
according to the $r$-band magnitude of the neighbour.
As shown in the legend, the median values in the four bins are 21.5, 23.0, 23.5 and 24.0. 
The mean bias and its uncertainty across all distance bins is indicated by the horizontal band.
}\label{fig:bias_vs_dgn}
\end{figure}

To explore the more practical question of how neighbours impact shear estimates
we divide the catalogue into bins according to neighbour distance.
Within each $d_{gn}$ bin, the galaxies are further split into twelve bins of input shear,
which are fitted to estimate the multiplicative and additive bias.
We show the result as the purple points in Fig. \ref{fig:bias_vs_dgn}, 
which can be compared with the earlier numerical model prediction in 
Fig. \ref{fig:dgn-vs-m-mc_toymodel}. 
The horizontal band on these axes shows the $1\sigma$ mean $m$ 
measured using all galaxies in the \hoopoe~ catalogue,
and sits at $m\sim -0.12$. 
We note a steeper decline than in the bold line (without the centroid cut), more akin to the case with the centroid cut ($\Delta r_0<1$ arcsec). 
This is not surprising given that the quality selection implemented by \imshape~includes exactly this cut. 
We do not report a local peak at $\sim 11$ pixels, which we saw before
in Fig. \ref{fig:dgn-vs-m-mc_toymodel}.
We suggested previously that effect was the result of positive $m$ in galaxies where the nearest neighbour is relatively faint and at middling distance. 
It is likely that many of these objects manifest themselves as large changes in 
other quantities to which \imshape's \blockfont{info\_flag} 
(see \citetalias{shearcat}) is sensitive
such as ellipticity magnitude and fit likelihood, 
or are flagged by the \se~ deblending cuts.

When divided into broad bins according to the $r$-band magnitude of the nearest neighbour $M_\mathrm{r,neigh}$
(the coloured stripes in Fig. \ref{fig:bias_vs_dgn})
we find the surviving objects show relatively weak dependence on neighbour brightness,
except at the neighbour distances, where bright neighbours have a slightly stronger (negative) impact than faint ones. 

We measure the additive bias components in the same bins,
but find no systematic variation with $d_{gn}$ above noise.

\begin{figure}
\includegraphics[width=1\columnwidth]{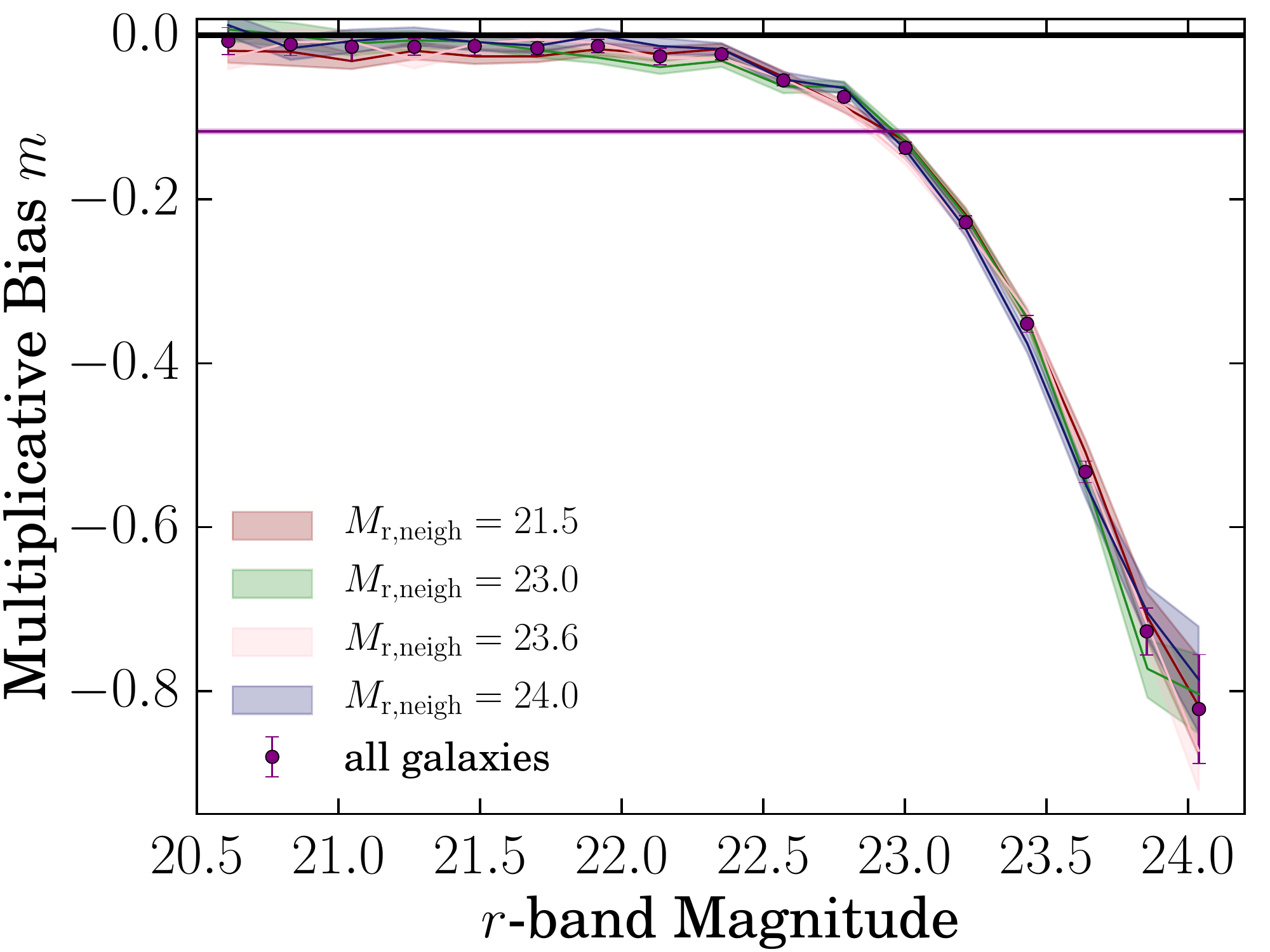}
\caption{Multiplicative bias as a function of $r$-band magnitude. 
As in Fig. \ref{fig:bias_vs_dgn} the four coloured bands represent equal number bins of neighbour magnitude. 
Purple points show the full catalogue, with no magnitude binning.
The mean bias and its uncertainty are shown by the purple horizontal band.
}\label{fig:m_vs_rneigh}
\end{figure}

Finally we show the analogous measurement in bins of galaxy magnitude in Fig. \ref{fig:m_vs_rneigh}. 
The steep inflation of $|m|$ at the faint end of this plot has been seen elsewhere (e.g. \citealt{shearcat, fc16}), 
and is easily understandable as the result of noise bias. 
We find that splitting by neighbour magnitude does not reveal any obvious trend here. 

\subsection{Untangling the Knot of Neighbour Bias}

\begin{figure*}
\begin{center}
\includegraphics[width=2\columnwidth]{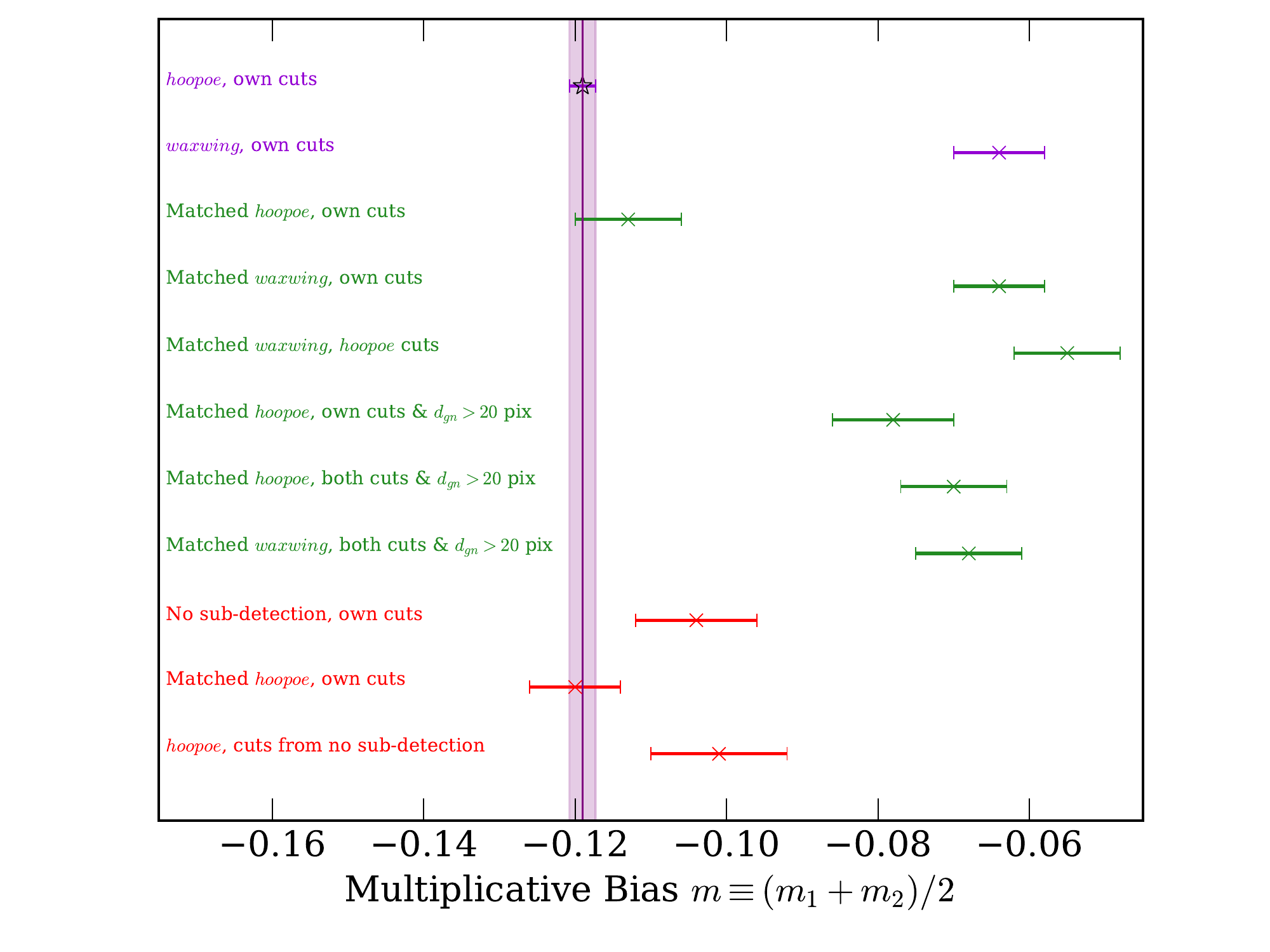}
\caption{Graphical illustration of the measured multiplicative bias in the various 
scenarios considered in this paper. 
The upper two lines show the mean $m$ in the main DES Y1 \hoopoe~simulations and a 
spin-off neighbour-free resimulation named \waxwing, as described in 
Section \ref{subsection:resims}. 
The middle section (green) shows results using only galaxies which appear in both 
the \hoopoe~and \waxwing~simulations. 
The matching process alone does not imply any quality-based selection function. 
The final three lines in red are from a similar matching between a smaller rerun of 
the simulation with and without sub-detection limit galaxies. 
See the text for details about each of these cases.
}\label{fig:m_summary_plot}
\end{center}
\end{figure*}

A central plank of this analysis rests on a comparison of the main 
\hoopoe~simulations with the neighbour-free \waxwing~resimulations
described in Section \ref{subsection:resims}.
The simplest comparison would be between multiplicative
bias values, calculated using all galaxies in each catalogue after cuts.
These values are shown by the two upper-most lines (purple) in
Fig. \ref{fig:m_summary_plot}.
The difference is an indicator of the net impact of neighbours through
any mechanism, which we find to be $\Delta m \sim -0.05$.

\begin{figure*}
\begin{center}
\includegraphics[width=1.04\columnwidth]{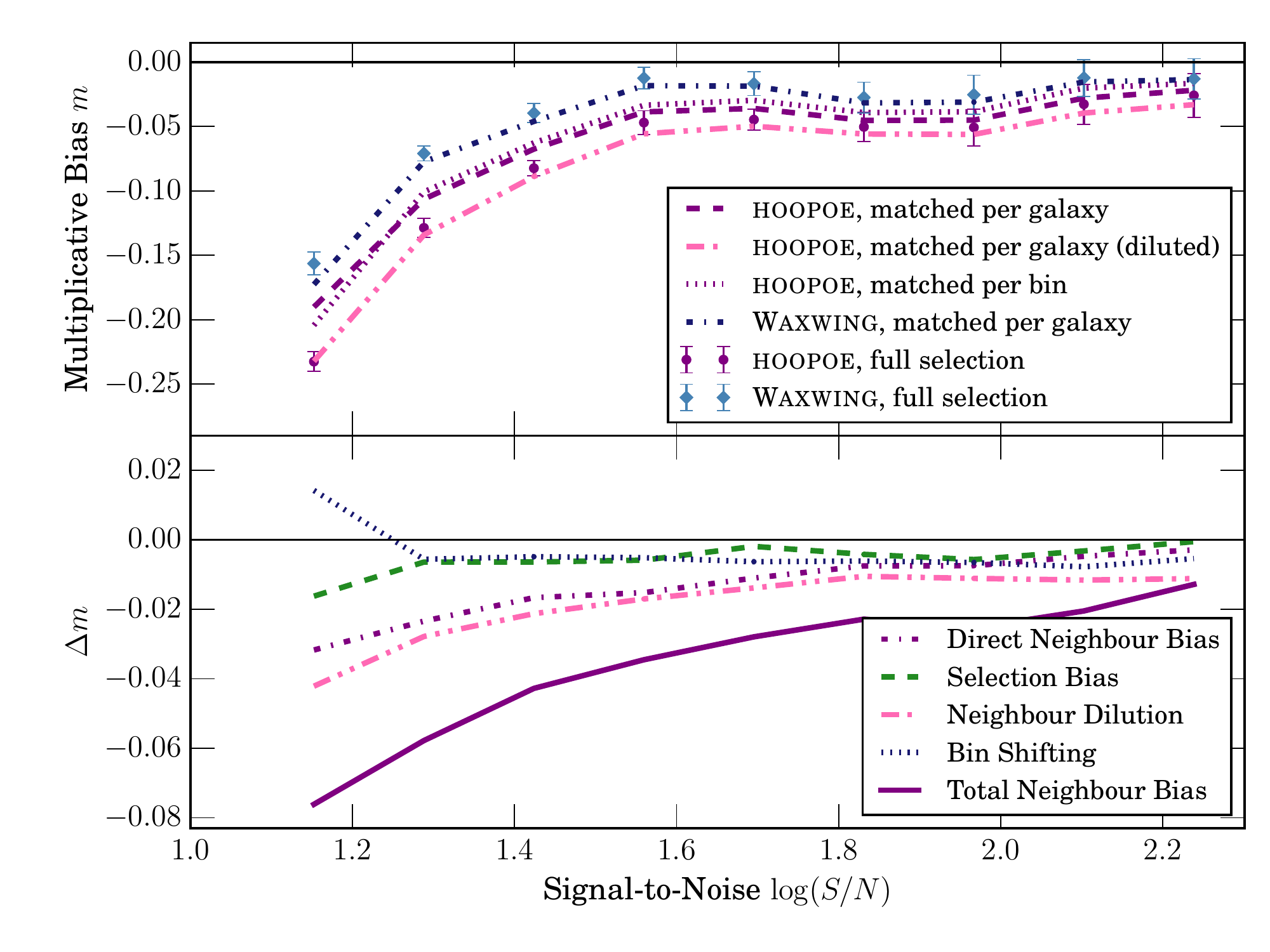}
\includegraphics[width=1.04\columnwidth]{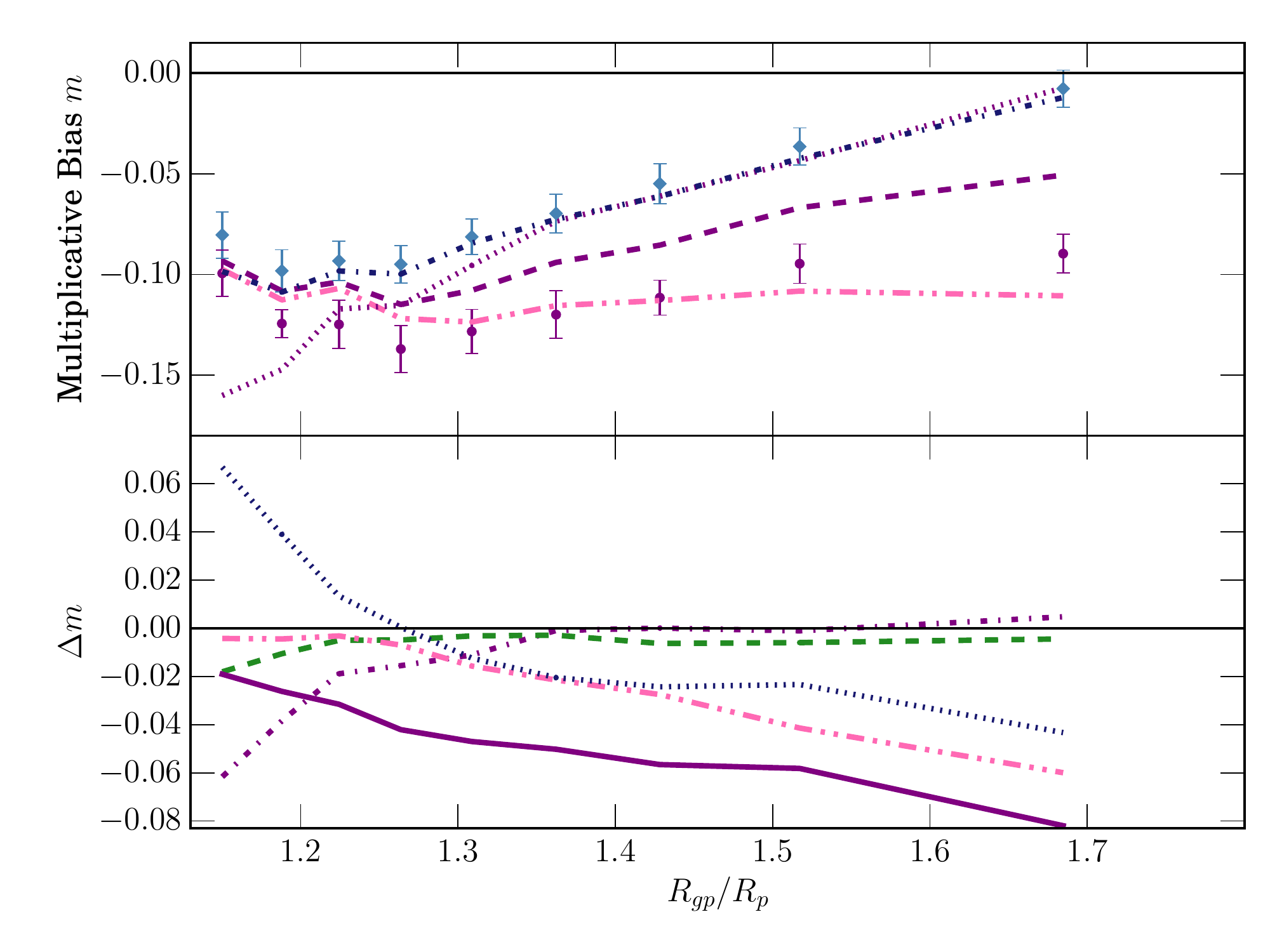}
\caption{\textbf{Top half of each panel} Multiplicative bias as a function of signal-to-noise and size. 
The purple circles show the measured bias using the main \hoopoe~simulation, 
and the blue diamonds are the resimulated neighbour-free version. 
The lines show permutations of the same measurements to
highlight the neighbour-induced effects causing the two to differ.
The dot-dashed blue and dashed purple lines show the impact of
applying the \hoopoe~selection mask to \waxwing~and vice versa.
The impact of bin shifting is shown by the purple dotted line,
which is calculated from the same matched galaxies, using the 
\hoopoe~shape measurements for the bias and \waxwing~size and 
\snr~for binning.
The pink curve is the same as the dashed purple, but with a fraction of
heavily blended galaxies added back with randomised shear (see Section \ref{sec:mechanisms:dilution}).
\textbf{Bottom half of each panel} The change in bias due to the effects described above.
The green (dashed) line shows the impact of selection effects
(the difference between the blue diamonds and the dashed line in the top panel).
The direct neighbour bias due to light contamination is shown by the purple dash-dotted line 
(purple dotted minus blue dash-dot top). 
The impact of shifting between bins is shown by the blue dotted (dashed minus purple dotted, top).
The pink dot-dot-dashed line illustrates the impact of adding back randomised shears, as described.
Finally the solid line represents the total neighbour bias, which includes all these effects 
(circles minus diamonds, top).
}\label{fig:money_plot_m}
\end{center}
\end{figure*}

To untangle the various contributions to this shift, we construct
a matched catalogue. Galaxies in the overlap between
\hoopoe~and \waxwing\ (12M galaxies over 183 square degrees)
are matched by ID; 
quality cuts are calculated for each set of measurments 
(see Appendix E from \citetalias{shearcat}).
Geometric masking from the DES Y1 \blockfont{Gold} catalogue
\citep{y1gold}
and \blockfont{SExtractor}
deblending flags are included for \hoopoe.
Since the latter are irrelevant to \waxwing, we omit them from quality flags on that dataset.
For conciseness we will refer to the two measurements as
``matched \hoopoe" and ``matched \waxwing",
and their cuts as ``\hoopoe~ cuts" and ``\waxwing\ cuts".
Since the images are identical in all respects, but for the presence of neighbours,
the statistical noise on the \emph{change} in measured quantities should be smaller than the 
face-value uncertainties.

The appropriate cuts are first applied to each catalogue, then the results are divided into equal number
signal-to-noise bins and fitted for the multiplicative bias in each.
The result is shown by the points in the upper left-hand part of Fig. \ref{fig:money_plot_m}.
The equivalent in bins of PSF-normalised size is shown on the right.
The difference between the blue and the purple points gives an indication of the total
effect of all neighbour-induced effects on $m$, indicated by the solid purple line in the lower panel. 
The generic shift attributed to ``neighbour bias" 
is in reality a collection of distinct effects.
By comparing the matched catalogues we identify four main mechanisms:
direct contamination, selection bias, $S/N$ bin shifting and neighbour dilution.
Each of these components that we describe is shown by one of the lines in 
Fig. \ref{fig:money_plot_m}.
For a visual summary of the various tests designed to isolate them see Fig. \ref{fig:m_summary_plot}.

\subsubsection{Direct Flux Contamination}
\label{sec:direct_bias}

The most intuitive form of neighbour bias arises from the 
fact that, even after masking, neighbours contribute some flux 
to the cutout image of a galaxy.
To gauge its impact we take the common sample of galaxies,
which pass cuts in both datasets.
The comparison is complicated somewhat by binning in measured $S/N$
or $R_{gp}/Rp$;
for this test, we divide both sets of galaxies using the \waxwing-derived
quantities.
The resulting $m$ measured using the \hoopoe~galaxies is 
unrealistic in the sense that we are binning measurements made in the presence of
neighbours by quantities derived from neighbour-subtracted images.
This exercise does, however, isolate the impact of the neighbour flux on 
the measured ellipticity.
The result is shown by the purple dotted and purple dot-dashed lines 
in the upper and lower panels of Fig. \ref{fig:money_plot_m}. 
The effect scales significantly with signal-to-noise and size.
Faint small galaxies are affected strongly by neighbour light,
while larger brighter ones are relatively immune.

\subsubsection{Neighbour-Induced Selection Bias}

To gauge the neighbour-induced selection effect,
we take the \waxwing~catalogue but now impose, in addition to its own quality cuts,
the selection function derived from the with-neighbour \hoopoe~dataset.
The double masking removes an additional 0.5M galaxies, which survive cuts in \waxwing\
but would be cut from the \hoopoe~catalogue.
The resulting change in $m$ is shown by the dot-dash blue lines
in the upper panels of Fig. \ref{fig:money_plot_m} (dashed green in the lower).
The multiplicative bias arising from this cut is less than one percent in all
but the faintest and smallest galaxies, where it can reach up to $m \sim -0.02$.

\subsubsection{Bin Shifting}

The above two tests encapsulate the impact on the measured ellipticities,
and the selection flags from neighbour flux.
An additional subtlety arises from the fact that
the measured quantities used to bin galaxies
(\snr~and \rgp)
are themselves affected by the presence of neighbours.
To test this we recalculate $m$ using the same galaxy selection
as in Section \ref{sec:direct_bias} (i.e. passing both sets of cuts),
but now binned by the appropriate \emph{measured} \snr. 
For clarity, the bin edges are unchanged,
defined to contain equal numbers of \waxwing~galaxies.
The result is shown by the dashed lines in Fig. \ref{fig:money_plot_m}.
The difference compared with the case using fixed binning is purely the result 
of galaxies moving between bins.
This shifting contributes multiplicative bias if one bins galaxies
by observed quantities such as \snr,
as we do in order to calibrate \imshape's shear estimates.
The amplitude of this is illustrated by the blue dotted line in the lower panels.
Such neighbour-induced shifting is noticable if we plot out the \snr\
of objects in \hoopoe~against the \snr~of the same objects in \waxwing.
Objects which are strongly shifted in \snr~are more likely to scatter upwards than downwards. 
A similar skew can be seen in the \rgp~plane; 
when galaxies are scattered in size they tend to be thrown further and more often 
upwards than downwards.
Small galaxies (which we know already are more strongly affected by noise bias)
are shifted strongly upwards by the presence of neighbour flux in the \hoopoe~images.
The result is a net negative $m$ added to the upper \rgp~bins,
and a simultaneous positive shift in the lowest bins from which galaxies are lost.
In the case of galaxy size we see the effects of bin scatter and direct neighbour bias almost negate each other,
although the degree of cancellation
is likely dependent on the specifics of the measurement code and the dataset. 

\subsubsection{Neighbour Dilution}\label{sec:mechanisms:dilution}

\noindent
A final point can be gleaned from Fig. \ref{fig:money_plot_m}: that applying the \waxwing~cuts to 
\hoopoe~induces a shift in $m$.
Naively one might expect the \hoopoe~selection function, which includes neighbours,
to remove the same galaxies as the \waxwing~selection, plus some extra strongly blended galaxies.
It is true that a sizeable number of galaxies are cut in the presence of neighbours, but would otherwise not be.  
There is also, however, a smaller population that survive cuts \emph{because} they have image plane neighbours.

We can see this clearly from the fact that the purple points and the dashed purple lines 
Fig. \ref{fig:money_plot_m} are non-identical. 
We identify three separate (but partially overlapping) galaxy selections in this figure:
(a) galaxies passing both sets of cuts,
(b) galaxies passing cuts in the absence of neighbours, but cut by the \hoopoe~selection
and
(c) galaxies which pass cuts in the presence of neighbours, but cut by the \waxwing\ selection. 
We find that populations (b) and (c)
have much smaller mean neighbour separation than the full population
(the histograms of $d_{gn}$ show a sharp peak at under 10 pixels).
In contrast, both the full catalogue and population (a) objects a much broader distribution 
($\bar{d}_{gn}\sim 24$ pixels). 

Based on the toy model predictions in Section \ref{section:toy_model}
we set out a working proposal:
that population (c), objects cut out only when neighbours are removed,
are extreme blends dominated by a superposed neighbour. 
We will assume these objects are boosted considerably in size, $S/N$ or both,
such that what would otherwise be a small faint galaxy is now sufficiently bright to pass quality cuts.
In these cases the measured shape of a simulated galaxy might be expected to be only weakly linked with the
input ellipticity. 
To approximate this effect we take population (a) \hoopoe~galaxies,
subject to both sets of cuts,
and add back some of the population (c) galaxies. 
Specifically, we include any objects
shifted in \snr~or \rgp~by more than $20\%$.
The true shears associated with these galaxies are now randomised
to eliminate any correlation with the measured ellipticity.
The result is shown as a pink dot-dot-dashed line in Fig. \ref{fig:money_plot_m}.
We can see that this effect, which
we call neighbour dilution, to good approximation accounts for the residual difference between the 
population (a) and (c) samples.
Particularly in the upper size bins of the right hand panel the differences are not eliminated entirely.
This is thought to be the result of residual (albeit weakened) covariance between the measured shapes of
strongly blended objects and the input shears.
Clearly the scenario in which a neighbour totally overrides the original shape of a galaxy is extreme,
and there will be an indeterminate number of moderate blends which are boosted sufficiently to survive
cuts but which retain some correlation with their original unblended shapes.
Such cases are, however, extremely difficult to model accurately with the resources available for 
this investigation.

\subsection{Isolating the Impact of Sub-detection Galaxies}

\noindent
A handful of previous studies have attempted to quantify the impact of galaxies below,
or close to, a survey's limiting magnitude.
For example, \citet{hoekstra15} and \citet{hoekstra16}  
suggest they can induce a non-trivial multiplicative bias, 
which is dependent on the exact detection limit.
They recommend using a shear calibration sample at least 
by 1.5 magnitudes deeper than the survey in question (which ours does).
Their findings, however, make exclusive use of the moments-based KSB algorithm
(see \citealt{kaiser95});
such techniques are known to probe a galaxy's ellipticity at larger radii than other methods,
which could in principle make them more sensitive to nearby faint galaxies.
It is thus a worthwhile exercise to to gauge their impact in our case with \imshape. 

\subsubsection{Impact on Multiplicative Bias}

We first compare
our \hoopoe\ simulations with the neighbour-free \waxwing~resimulations.
Since \waxwing~postage stamps consist of only a single profile
added to Gaussian pixel noise, 
they are unaffected by neighbours of any sort
(faint or otherwise).
We have seen that the impact of neighbours is strongly localised,
with the excess $m$ converging within a nearest neighbour distance $d_{gn}$
of a dozen pixels or so.
Thus selecting galaxies that are well separated from their nearest
\emph{visible} neighbour will isolate the impact of the
\emph{undetected} ones.

A further cut is thus imposed on $d_{gn}<20$ pixels. 
Relative to the case with quality cuts only,
the global multiplicative bias now shifts from 
$m\sim -0.119$
to 
$m=-0.064\pm0.006$
(the first and second lines in green on Fig. \ref{fig:m_summary_plot}).
This measurement is in mild tension with the value measured from \waxwing\ 
(again under its own cuts, with the selection on $d_{gn}$).
This difference, which amounts to a negative shift in $m$ of $\sim 0.01$ is,
we suggest, the net effect of the sub-detection galaxies.
From these numbers alone we cannot tell if this is a result of selection effects,
flux contamination, bin shifting or some combination thereof. 

Interestingly we find that imposing both the \hoopoe~and \waxwing~selection functions
in addition to the cut on $d_{gn}$ 
brings $m$ into agreement to well within the level of statistical precison 
(compare the final and penultimate lines in green in Fig. \ref{fig:m_summary_plot}). 
That is, when restricted to a subset of galaxies that pass quality cuts in both simulations
the flux contributed by the faint objects has little impact.
Their main impact is rather that they allow a population of marginal faint galaxies 
which would otherwise be flagged and removed by quality cuts
to pass into the final \hoopoe~catalogue.

\begin{figure}
\includegraphics[width=\columnwidth]{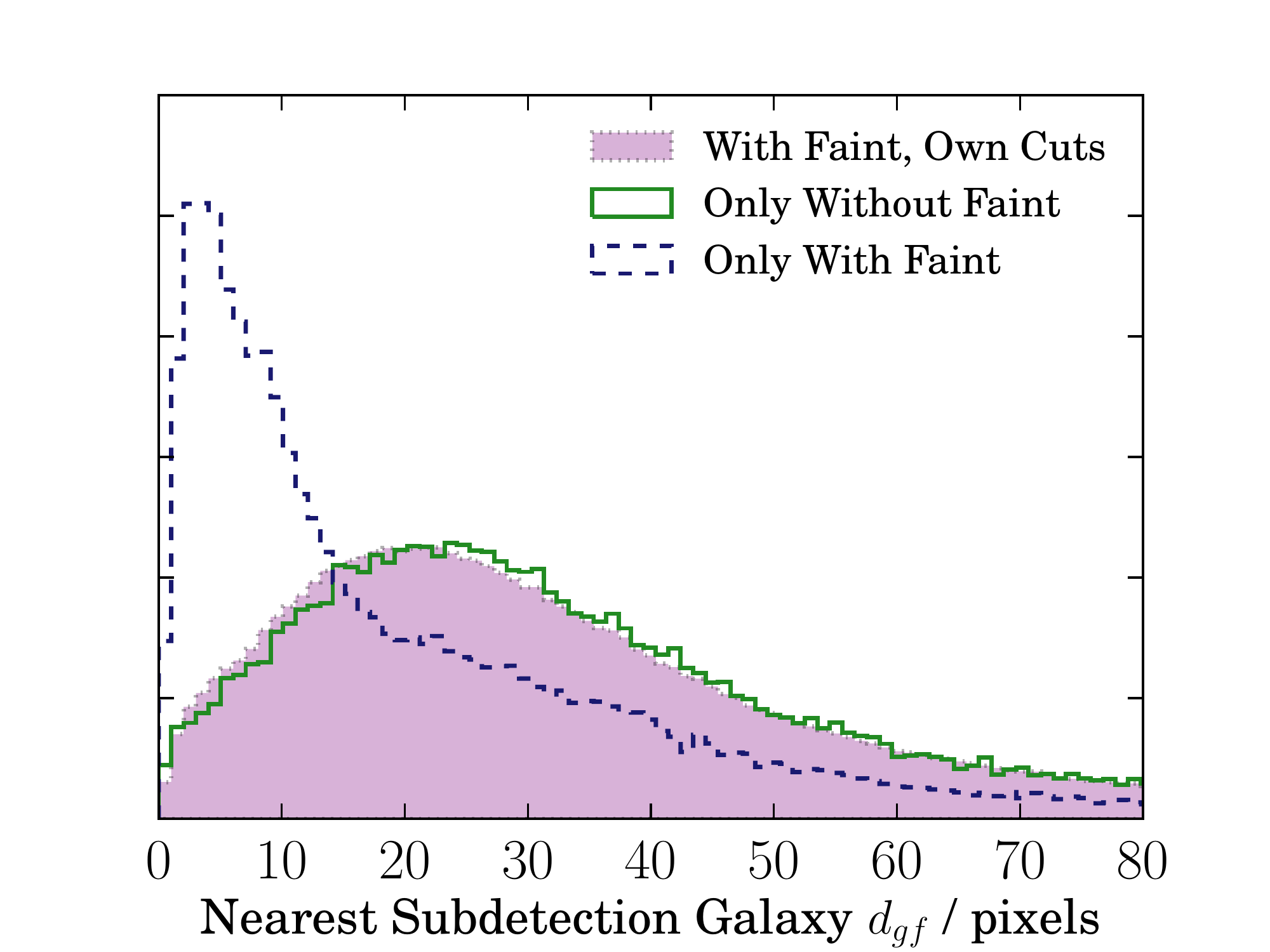}
\caption{Histogram of radial distances between galaxies in our measured shape catalogues (the full \hoopoe~simulations) 
and the nearest object below the DES detection limit.
The dotted line includes all objects prior to quality cuts, while the solid line shows the impact of applying 
\imshape's \blockfont{info\_flag} cuts (see \citetalias{jarvis16}).
The dashed blue line shows the population of galaxies which survive cuts only in the presence of the faint galaxies.
}\label{fig:faint_distance_hist}
\end{figure}

To test this idea further we rerun a subset of 100 random tiles from the simulated footprint, 
without the final step of adding sub-detection galaxies. 
To minimise the statistical noise in this comparison we enforce the same COSMOS profiles, shears and rotations 
as well as the per-pixel noise realisation as before.
\se~source detection is applied and the blending flags are propagated into the postprocessing cuts.

The raw $m$ values calculated from the rerun and the main \hoopoe\ simulations, 
matched to the same tiles, are shown by the upper two red lines on 
Fig. \ref{fig:m_summary_plot}.
The downward shift of $\sim0.01$ is consistent with the previous result
based on the main simulation.
This comparison should encapsulate the full effect of the faint objects
(since there are no other differences between these datasets).

For each galaxy we next measure the distance to the nearest faint object $d_{gf}$,
the distribution of which is shown under various selections in
Fig. \ref{fig:faint_distance_hist}.
Like in the comparison in Section \ref{sec:mechanisms:dilution}, there 
is a population of galaxies that survive cuts \emph{only}
in the simulation with the sub-detection objects,
and these galaxies tend to be ones with extremely close faint neighbours.
Interestingly the inverse population
surviving only when they are removed 
do not preferentially have small $d_{gf}$.
This is intuitively understandable:
a faint object might boost its neighbour's apparent size or \snr\
if it were centred within a few pixels.
Otherwise it would act as a source of background noise, which would
reduce the quality of the fit.

Finally we find that if we apply both selection functions to the with-faint galaxies, 
the measured biases become roughly consistent.
These findings, combined with the observations in the previous section lead us to an interesting conclusion:
the major effect of the faint galaxies in the DES Y1 \imshape~catalogue
is to allow a population of small faint galaxies
to pass quality cuts, where otherwise they would have been removed.  
This is analogous to the neighbour dilution effect described above,
but is subdominant to the influence of visible neighbours.

\subsubsection{Impact on Background Flux Subtraction}

As a test of the robustness of this result we recompute our \imshape\
fits on the faint-free images,
with and without the correction for the shift in the background flux that
\emph{would} have been applied if the sub-detection galaxies had been drawn. 
The mean per-tile correction is $\Delta f \sim 0.05$,
against typical noise fluctuations $\sigma_n \sim 6.5$.
Matching galaxies and examining the histograms of 
$\Delta S/R$ and 
$\Delta R_{gp}/R_p$
reveals weak downwards scatter in both quantities
(i.e. the flux subtraction alone makes galaxies appear smaller and fainter).
The magnitude of the shift is, however, tiny, peaking at $\sim -0.1$ and $-0.005$ respectively.
This is logical given the definition in equation \ref{equation:snr_definition}.
If the change is small enough such that the best-fitting model is stable,
then an incremental reduction in flux will reduce the signal-to-noise of the measurement. 
Looking at the best-fit shapes, we find a small shift towards high ellipticities,
which can likewise be understood as a numerical effect;
imposing a flat positive field of zero ellipticity will dilute the measured shear,
producing a bias towards round $|e|$.
The reverse logic applies with the flux correction,
and subtracting a flat value from all pixels will make galaxies appear
slightly \emph{more} elliptical.
In practice we find a sharp peak at $\Delta e \sim 0.001$.

\subsection{Suppressing Neighbour Bias}

There is no universal definition for the shape-weighted effective number density
commonly used as proxy for cosmological constraining power in a shear catalogue.
One which is particularly useful in the context of weak lensing,
and which has been adpoted in DES Y1
is the prescription of \citet{chang13}, which is designed to account for shape noise
and fitting error (see equation 7.5 in \citealt{shearcat}).
A second useful definition is set out by \citet{heymans12} in terms of the
(see also \citetalias{shearcat}). 
We compute a neighbour distance $d_{gn}$ for every object in the real data,
which allows us to cut on this quantity.
Removing any galaxy with a neighbour detected within a radius of $d_{gn}=20$ pixels
reduces the effective number density of sources using either definition 
to about $70\%$ of its initial value,
from
$n^\mathrm{H13}_\mathrm{eff}=5.48$ to 
$n^\mathrm{H13}_\mathrm{eff}=3.68$ arcmin$^{-2}$
using \citet{heymans12}'s definition.
Using the prescription of
\citet{chang13}, 
the equivalent density drops from
$n^\mathrm{C13}_\mathrm{eff}=3.18$ prior to cuts and 
$n^\mathrm{C13}_\mathrm{eff}=2.18$ arcmin$^{-2}$
afterwards.
This cut is stringent,
as we have shown that beyond $\sim 12$ pixels the multiplicative bias becomes
insensitive to further selection on $d_{gn}$.
There are, however, a number of limitations in our analysis,
including the fact that $d_{gn}$ is defined using the true input positions,
and indeed that we are using only the detected positions in DES to draw our simulated $M_\mathrm{r}<24.1$ galaxies.
We thus judge that a level of conservatism is appropriate here.
Relaxing the cut to $d_{gn}>14$ pixels leaves $n_\mathrm{eff}$ at $84\%$ of its full value.


\section{Cosmological Implications}
\label{section:cosmology}

As we have shown in the previous sections, if ignored completely image plane neighbours 
can induce negative calibration biases 
in \imshape\
of a few per cent or more. 
The earlier part of the investigation focused on \emph{when} and \emph{how} 
neighbour bias can arise, first in the context of single-galaxies 
and then on ensemble shear estimates. 
We now turn to a more pressing question from the general cosmologist's perspective: 
\emph{how far should I be concerned about these effects in practice}?
We present a set of numerical forecasts using 
the \blockfont{MultiNest} nested sampling algorithm \citep{multinest}
to sample trial cosmologies.
Each of the likelihood analyses presented in this paper
has been repeated using a Markov Chain Monte Carlo sampler
(\blockfont{emcee}).
Although we see the same small shift in contour size noted by \citet{keypaper}
(see their Appendix A),
which diminishes as the length of the MCMC chains increases, 
we find our conclusions are robust to the choice of sampler.  
Our basic methodology here follows previous numerical forecasts 
(e.g. \citealt{jb10,krause16,y1methodology}).
We construct mock DES Y1 cosmic shear measurements using a matter power spectum derived 
from the Boltzmann code \blockfont{CAMB}\footnote{http://camb.info}
with late-time modifications from \blockfont{halofit}.
The cosmic shear likelihood surface is sampled at trial cosmologies using 
\blockfont{CosmoSIS}\footnote{https://bitbucket.org/joezuntz/cosmosis}. 
The final data used for the likelihood calculation have the form of real-space $\xi_\pm$ 
correlations.
For the photometric redshift distributions we use the
measured estimates in four tomographic bins,
obtained from runs of the 
\blockfont{bpz} code on the Y1 \imshape~catalogue,
as described by \citet{photoz}.
Since this analysis was completed before the details of 
the photometric redshift calculation for DES Y1 had been finalised,
these distributions
differ marginally from (but are qualitatively the same as) the final
version used in \citet{shearcorr} and \citet{keypaper}.
In all chains which follow we maginalise over two nuisance parameters 
(an amplitude and a power-law in redshift) for 
intrinsic alignments, photo-$z$ bias and shear calibration bias. 
In total this gives 10 extra free parameters in addition to six 
for cosmology (\Omegam, $\Omega_\mathrm{b}$, \as, \ns, $h$, $\Omega_{\nu}h^2$),
which are also allowed to vary. 
Apart from the difference in redshift distributions remarked upon above,
our analysis choices match the DES Y1 cosmic shear analysis of \citet{shearcorr}.
We refer the reader to that paper for details of the priors and scale cuts,
and their derivation. 
Finally, the following adopts shear-shear covariance matrices derived from the analytic halo model calculations of 
\citet{methodpaper}.
We assume a fiducial $\Lambda$CDM cosmology 
$\sigma_8=0.83$, \Omegam $=0.295$, $\Omega_\mathrm{b}=0.047$, \ns $=0.97$, $h=0.688$, $\tau=0.08$, 
with non-zero comoving neutrino density $\Omega_{\nu}h^{2}=0.00062$. 

\subsection{Mean Multiplicative Bias}\label{subsection:cosmology:scale_independent_m}

\begin{figure*}
\begin{center}
\includegraphics[width=\columnwidth]{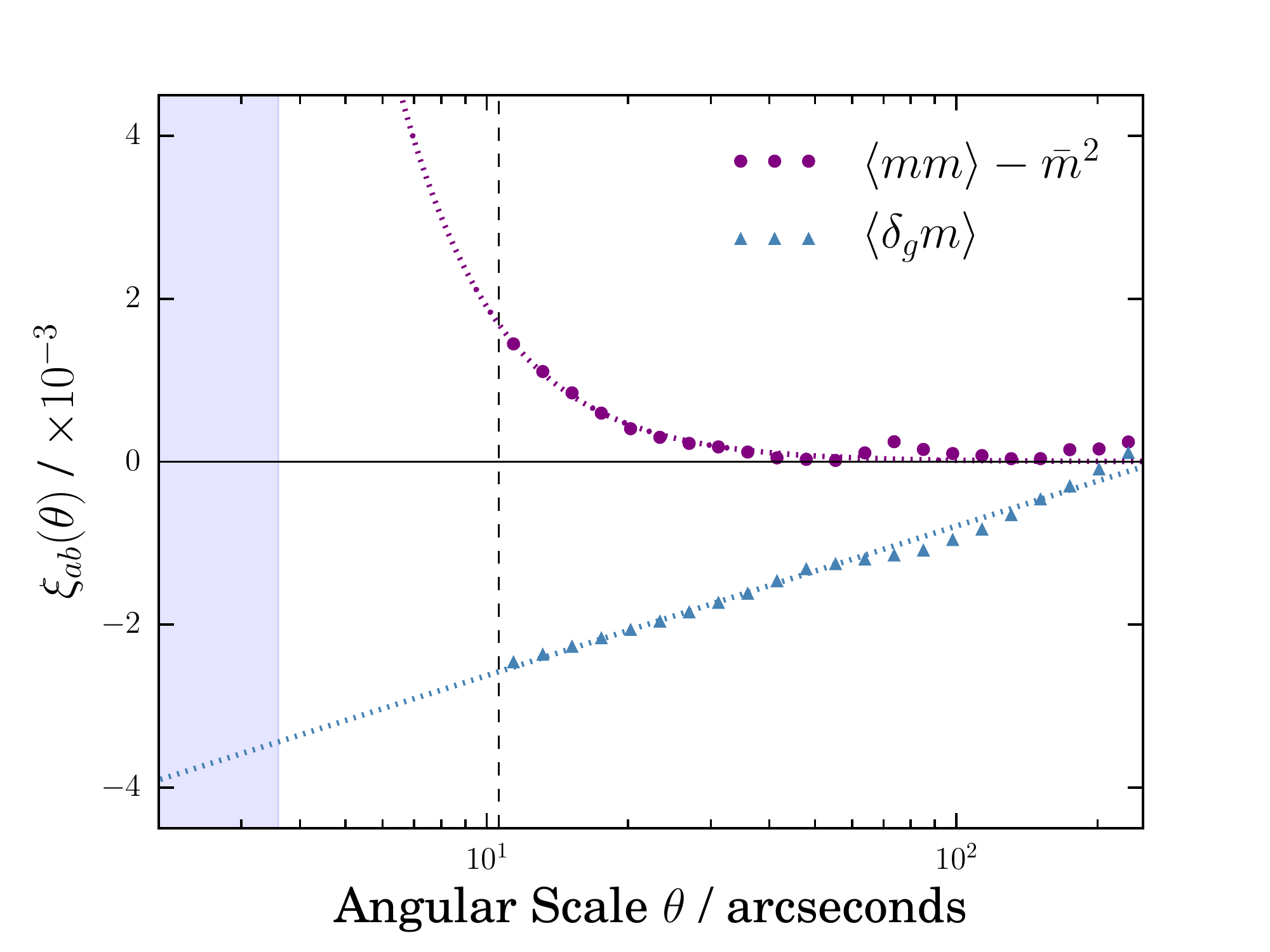}
\includegraphics[width=1.75\columnwidth]{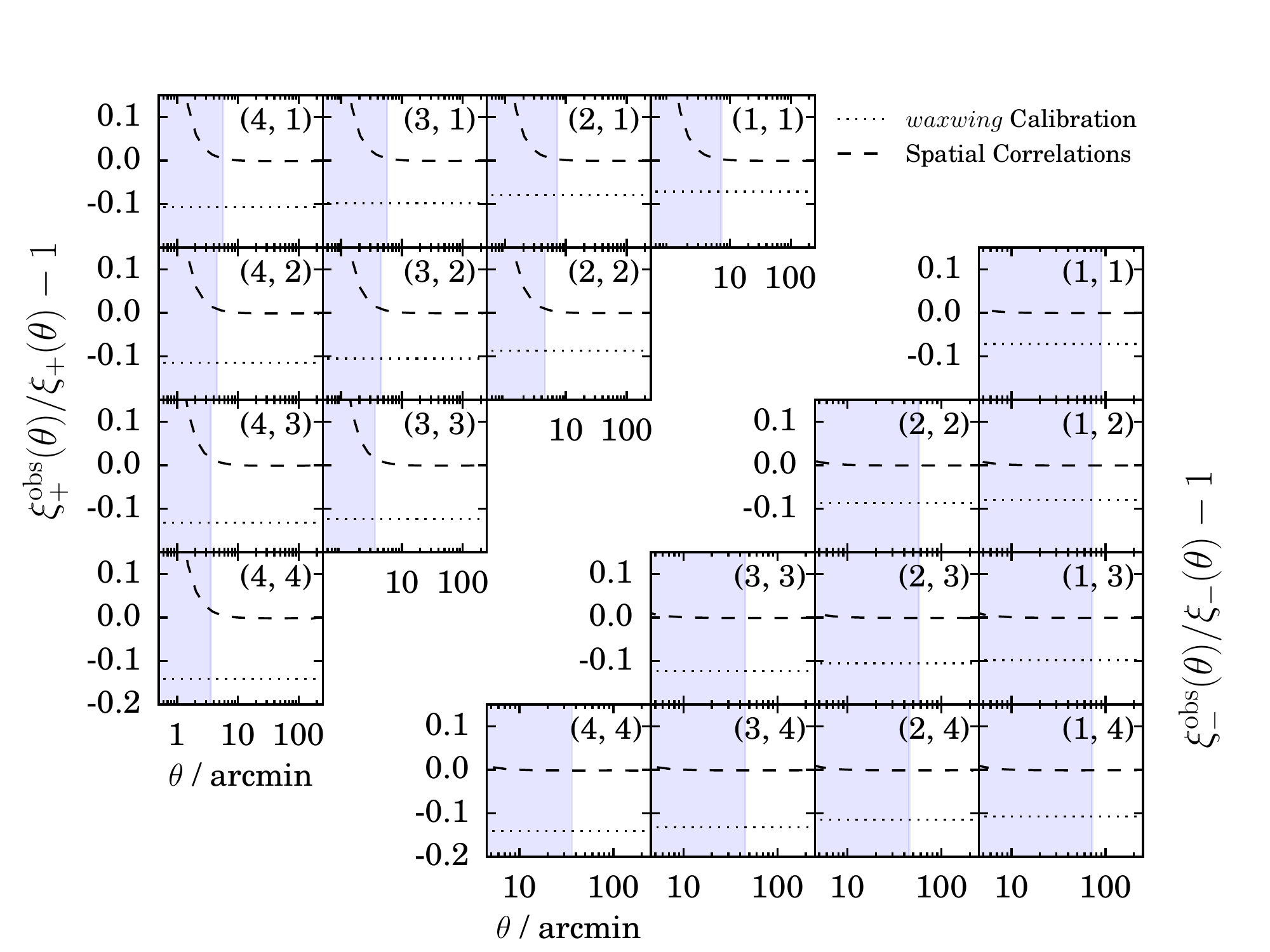}
\caption{\textbf{Top}: The observed two point correlation of multiplicative bias,
as measured from the main \hoopoe~simulation set presented in this paper.
Sub-patches are used to compute $m$ in spatial patches of dimension $0.15 \times 0.15$
degrees and the correlation function calculated as described in the text.
The dashed vertical line shows the diagonal scale of the sub-patches, below which we do
not attempt to directly measure spatial correlations.
The shaded blue bands show the minimum and maximum scales used in the DES Y1 cosmic shear
analysis of \citet{shearcorr}. 
\textbf{Bottom}: Residuals between the mock two point shear-shear data used in this paper, 
before and after different forms of bias have been applied.
The upper-left and lower-right triangles show the $\xi_+$ and $\xi_-$ correlations
respectively, calculated using the redshift distributions of \citet{photoz}.
The dotted black lines, which are flat across all scales but vary between panels,
show the result of calibrating our Y1 shear measurements with a simple postage stamp 
simulation without image plane neighbours. 
The dashed lines illustrate the impact of ignoring scale-dependent selection effects, 
which are not captured by our simulation-based calibration.
The shaded blue regions of each panel show the excluded scales for each particular 
tomographic bin pairing.
}\label{fig:xi_cornerplot}
\end{center}
\end{figure*}

We first seek to quantify the bias that would be present in a cosmic shear analysis in 
a survey like DES, if we were to use a simple postage stamp simulation of the sort 
presented in \citetalias{jarvis16} and \citet{miller13}. 
To this end we use the neighbour-free \waxwing~dataset to construct an alternative shear calibration.
In \citetalias{shearcat} we compared three methods for shear calibration using the \hoopoe~ 
simulations and found our results to be robust to the differences.
We now use the fiducial (grid-based) scheme to derive an 
alternative set of bias corrections from \waxwing.
These are then applied to the same galaxies in the matched \hoopoe~simulation, 
and residual biases are measured in four DES-like tomographic bins.
The process is very similar to the diagnostic tests in \S 5 of \citetalias{shearcat}, 
and so we defer to that work for details of the redshift bin assignment of simulated galaxies.

Using the neighbour-free simulation we under-correct the measurement bias by several percent in each bin.
The remeasured residual bias after calibration provides an estimate for the level of systematic 
that would be present were we to calibrate DES Y1 using the simpler \waxwing~ simulations. 
In the four tomographic bins used in DES Y1
we find $(\Delta m^{(1)}, \Delta m^{(2)}, \Delta m^{(3)}, \Delta m^{(4)}) = (-0.037, -0.044, -0.064, -0.073)$,
and apply these biases to our mock data. 
The resulting shift in the shear two-point correlations is shown by the black dotted lines in the 
lower panel of Fig. \ref{fig:xi_cornerplot}.
Since the calibration scheme does not explicitly include neighbour distances, but rather orders galaxies 
into cells of $S/N$ and $R_{gp}/R_p$, 
this test does not include any scale dependent neighbour effects. 
The calibration effectively marginalises out $d_{gn}$, and the residual biases are an average over the survey. 
For the moment we will assume this mean shift in $m$ is sufficient, and return  to the question of scale dependence in 
the following section. 

\begin{figure}
\begin{center}
\includegraphics[width=\columnwidth]{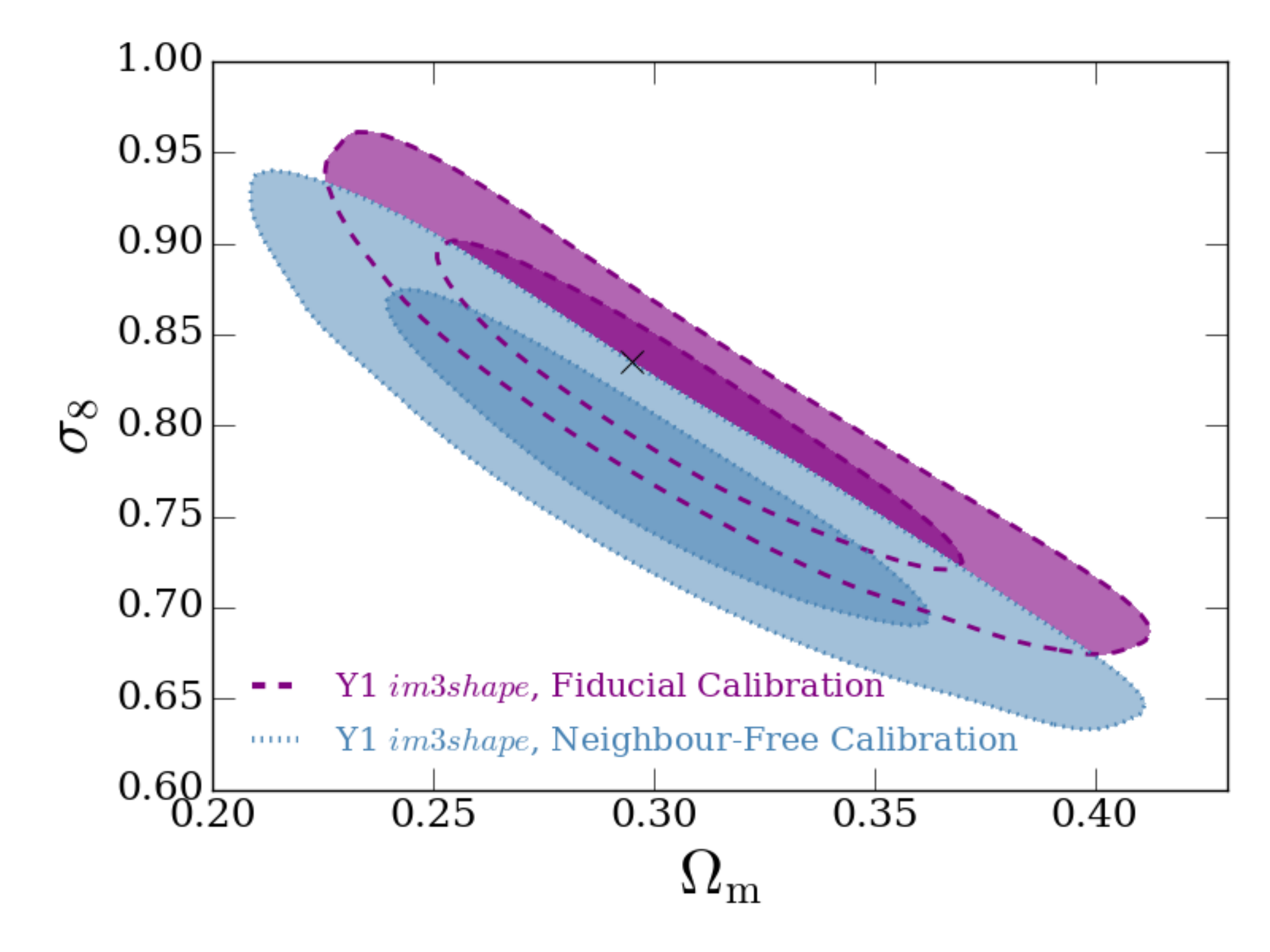}
\caption{Expected cosmology constraints from DES Y1 cosmic shear only. 
The purple (solid) contours show the results of calibrating using a simulation which fully 
encapsulates all biases in the data, leaving no residual $m$ in the final catalogue. 
In blue (dash-dotted) we show the result of calibrating with an insufficiently realistic simulation, 
which leaves a residual bias between $-0.03$ and $-0.08$ in each of the redshift bins. 
For reference we mark the input cosmology with a black cross.  
}\label{fig:cosmology_money_plot}
\end{center}
\end{figure}

Our predicted cosmology constraints with weak lensing alone in DES Y1 are shown in Fig. \ref{fig:cosmology_money_plot}.
In purple we show the results of the fiducial analysis, 
in which the shear calibration fully captures all neighbour effects and leaves no residual multiplicative bias. 
The blue (solid) contours then show the impact of residual neighbour biases per bin at the level described.
As we can see, even when marginalising over $m^i$ with an (erroneously) zero-centred Gaussian prior of width 
$\sigma_m=0.035$,
our cosmology constraints are shifted enough to place the input 
cosmology outside the $1\sigma$ confidence bounds.
We reiterate here that this calculation highlights the bias that 
\emph{would} arise were we to naively apply a calibration
of the sort used in DES SV based on neighbour-free simulations to the Y1 data.
Since we are confident that the \hoopoe~code captures the effects of image plane neighbours correctly
(at least to first order) this is a hypothetical scenario only and not a prediction of actual bias in DES Y1. 

\subsection{Scale Dependence}
\label{subsection:cosmology:scale_dependent_m}

It is not trivial that including an mean neighbour-induced component
to $m$ over the entire survey will be sufficient to mitigate all forms of neighbour bias.
The local mean $m$ on a patch of sky is sensitive to spatial fluctuations in source density,
which could induce scale dependent bias on arcminute scales.
Clearly, when correlating galaxy pairs on small scales one can expect a larger fraction in which
the objects come from a similar image plane environment,
and more often than not that enviroment will be densely populated. 
Thus the true multiplicative bias should become more negative on small scales. 

\begin{figure}
\begin{center}
\includegraphics[width=\columnwidth]{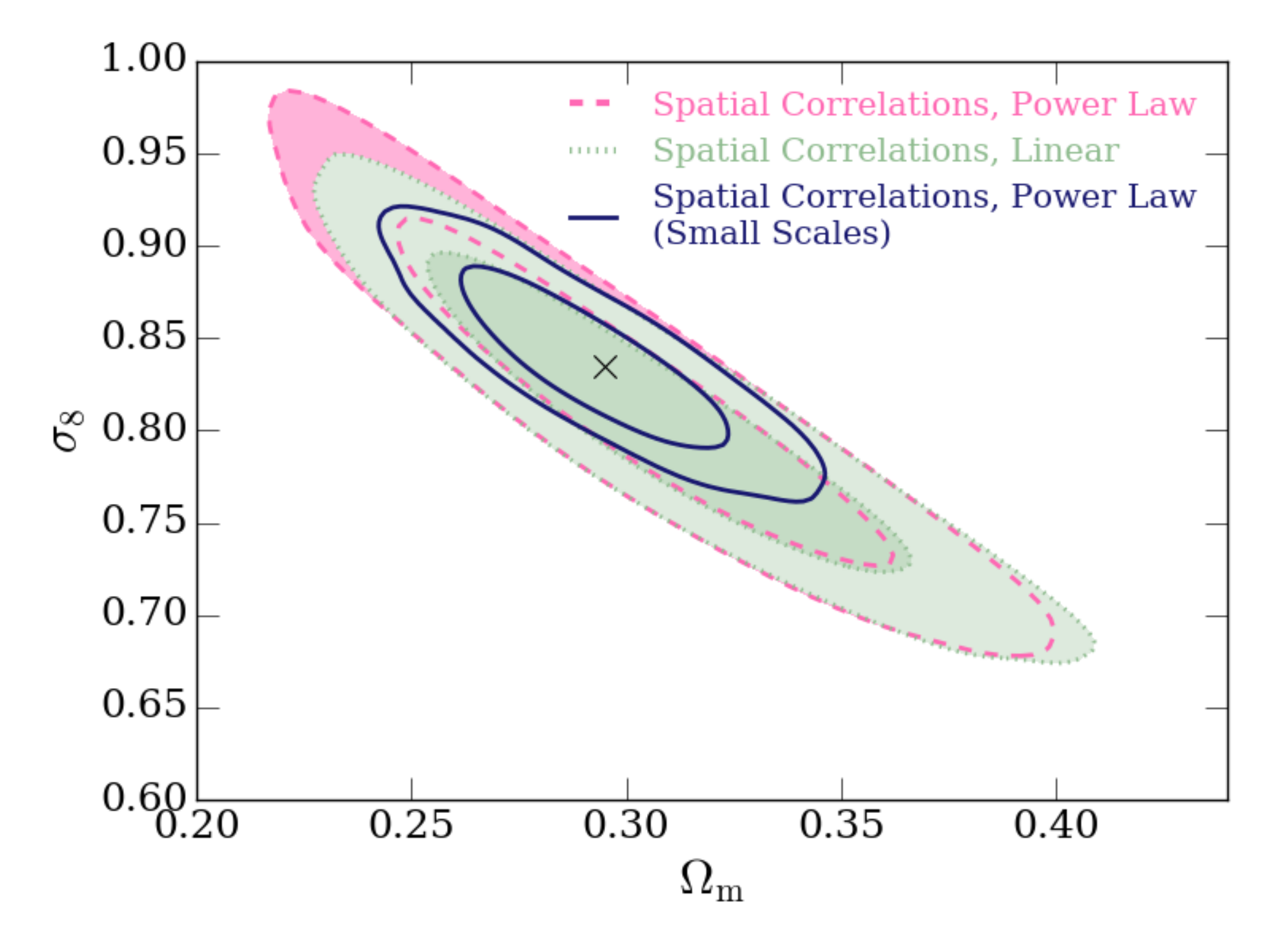}
\caption{The same as Fig. \ref{fig:cosmology_money_plot}, but now showing the impact of 
residual scale dependent selection bias.
The two sets of confidence contours represent different assumptions about
the small scale extrapolation of the $\xi_{mm}$ correlation, as
outlined in the Section \ref{subsection:cosmology:scale_dependent_m}.
In green (dashed) we show a mildly optimistic case, using the linear fit
shown in Fig. \ref{fig:xi_cornerplot}.
The pink dotted contours show a (strongly pessimistic)
power law extrapolation.
The dark blue solid line makes identical assumtions to the pink, but incorporates
small-scale information, to a minimum separation of 
$\theta^{+}_\mathrm{min}=0.5$ arcminutes in $\xi_+(\theta)$
and 
$\theta^{-}_\mathrm{min}=4.2$ arcminutes in $\xi_-(\theta)$.
As in Fig. \ref{fig:cosmology_money_plot} the input cosmology is shown by a black cross.
}\label{fig:cosmology_ximm}
\end{center}
\end{figure}

Two subtly different effects emerge from this thought experiment.
First, the multiplicative bias of galaxies will be spatially correlated 
i.e. a correlation involving two galaxy populations 
$\left<m^im^j\right>$ 
is not just the product of the means $\bar{m}^i\bar{m}^j$.
Second, in the small $\theta$ bins one is selecting galaxies with close partners with
which to correlate, and thus oversampling the dense parts of the image.   
To gauge the level of these effects, we divide each simulated coadd tile into a grid of 25 square
sub-patches with dimension 
$0.15\times0.15$ degrees. 
We fit for $m$ using the galaxies in each sub-patch and assign the resulting value to these objects. 
While this only allows a noisy measurement of $m$, it should capture the spatial variations in 
number density to the level of a few percent.
We next measure the two-point correlation function of multiplicative bias values assigned in this way
using \blockfont{TreeCorr}\footnote{http://rmjarvis.github.io/TreeCorr},
excluding galaxy pairs at angular separation smaller than the scale of the sub-patches. 
We refer to this bias-bias autocorrelation as $\xi_{mm}$, 
which we show as a function of angular scale in the upper panel of Fig. \ref{fig:xi_cornerplot}.
analogously one could use the sub-patches to construct correlations between $m$ and galaxy number density
$\xi_{gm}$
or density with density
$\xi_{gg}$.
The statistical noise on these correlations
is significantly lower than that on the individual sub-patches
by virtue of the large simulation footprint.
Note that in Fig. \ref{fig:xi_cornerplot} we subtract 
$\bar{m}^2$, measured from all galaxies in the simulation,
from the measured $\xi_{mm}$.
If there were no $\theta$ dependence the correlation 
$\langle m^i m^j \rangle$ should simply average to the square of the global mean in all scale bins. 
As we can see from the circular points in this figure, scales larger than 
the diagonal size the sub-patches 
(shown by the vertical dashed line) exhibit non-negligible excess $\xi_{mm}$.
One obvious question is whether this could be the result of finite binning error, 
which scatters galaxies in the same sub-patch into different $\theta$ bins.
To verify this is not the case we repeat the measurement as before, but
halve the parameter controlling binning error tolerance (``bin slop") and obtain the same results.

To extend this measurement down to scales below the sub-patch size we must make some assumptions about the functional
form of the $mm$ correlation. We fit a power law, 
$ \Delta \xi_{mm}(\theta) \equiv \xi_{mm}(\theta) - \bar{m}^2$:

\begin{equation}
\Delta \xi_{mm}(\theta) = \beta \theta^{-\alpha},
\end{equation} 

\noindent
which is shown by the dotted purple line in this figure.
This provides a qualitiatively good fit to the measured points,
but as we can see implies a rather dramatic inflation on small scales.

In the limited range over which we have a nonzero measured correlation, however,
a linear function of $\theta$ (truncated at $\theta = 27$ arcminutes)
also provides a reasonable by-eye fit.
The small-scale extrapolation in this case is significantly milder.
The $\left< m \delta_g\right>$ and $\left< \delta_g \delta_g\right>$
measurements are linear with $\theta$ to good approximation,
and so we use linear fits to extrapolate them below
the patch size.

Assuming the bias per tomographic bin can be written as the sum of a
redshift dependent contribution (i.e. a scale invariant mean in each bin), 
and a scale dependent term, one can write the correlation per bin as 
$m^{i} m^{j}=\bar{m}^i\bar{m}^j+ \Delta \xi_{mm}(\theta)$.
A more complete derivation of this expression can be found in 
Appendix \ref{appendix:2pt_m}.
The first part can be extracted from the DES Y1 calibration, 
and we can fit for $\Delta \xi_{mm}(\theta)$ as described above.
A set of modified $\xi^{ij}_{\pm}$ are thus computed.
These appear in the lower panels of Fig. \ref{fig:xi_cornerplot} as dashed lines. 
As we can see, the scale cuts of \citet{shearcorr} 
(excluded scales are shaded in blue) are sufficiently stringent to remove almost all 
of the visible scale dependence. 
Though reassuring for the immediate prospects of DES Y1, this will not trivially be true 
for all future (or indeed ongoing) lensing surveys.
It is thus important that the effects we identify here are properly understood at a level beyond the
resources of the current paper. 
These biased data are then passed through our likelihood pipeline to gauge the cosmological impact,
which is shown in Fig. \ref{fig:cosmology_ximm}.
In the linear case (dashed green) there is no discernable bias in the 
$\sigma_8$ \Omegam~pair;
even the much harsher power-law extrapolation (pink dotted) 
induces only an incremental shift along the degeneracy direction.
In both cases the input cosmology still sits comfortably 
within the $1\sigma$ confidence contour.

Finally we test the impact of relaxing the stringency of our scale cuts. 
The minimum scales used
for $\xi^{ij}_+$ and $\xi^{ij}_-$ are shifted downwards to
$0.5$ and $4.2$ arcminutes respectively, irrespective of bin pair,
which are the cut-off values used in fiducial cosmic shear analysis of \citet{hildebrandt16}. 
This increases the size of our datavectors considerably.
Incorporating smaller angular scales will clearly improve the constraining power of the data to an extent.
Primarily the effect is to shorten the lensing degeneracy ellipse, cutting out much of the peripheral curvature,
but it also reduces the width in the $S_8$ direction.
These scales, however, contain biased information, which induces tension between the small and large angular scales.
With the strongest (power law) scale dependence considered, the input cosmology is displaced marginally along the degeneracy curve,
though it remains comfortably within the $1\sigma$ confidence bound.


\section{Conclusions}\label{section:conclusion}

The Dark Energy Survey is the current state of the art in cosmological weak lensing.
Multi-band imaging down to 24th magnitude across 1500 square degrees of
the southern sky has yielded hitherto unparalleled late-time constaints on
the basic parameters of the Universe
(see \citealt{shearcorr} and \citealt{keypaper}). 

In this paper we have used one of two DES Y1 shear catalogues, 
and large-area simulations based upon them,
to quantify the impact of image plane neighbours
on both ensemble shear measurements, and on the inferred cosmological
parameters.

In order to properly mitigate the influence of galaxy 
neighbours, and thus avoid drawing flawed conclusions about cosmology from
the data,
it is important to first understand the mechanisms by which they enter
the shape measurement.
Using a simple toy model of the galaxy-neighbour system we have shown that 
shear bias can arise even when the distribution of neighbours
is isotropic
(i.e. there is no preferred direction).
This is the result of a small difference in the impact of the same neighbour,
when it is placed on or away from the axis of the shear.
We have furthermore shown that the resulting multiplicative shear bias $m$ can
be either positive or negative, 
depending on the model parameters.
With slight modifications to the toy model, 
whereby we Monte Carlo sample input parameters from the joint distribution 
of the equivalent properties measured in DES Y1,
we have shown that a mild negative $m$ is dominant
when marginalising over a realistic ensemble of neighbours.
This was seen to be strongly dependent on the distance of the neighbour,
and to be mitigated but not eliminated by basic cuts on the centroid position 
of the best-fitting model.

Using the DES Y1 \hoopoe~simulations, 
which were also used to derive shear calibration corrections for the Y1
\imshape~catalogue of \citetalias{shearcat},
we have presented a detailed study of the ensemble effects of galaxy neighbours.
In this analysis we have identified four mechanisms for neighbour bias,
which we call 
flux contamination, selection effects, bin shifting and neighbour diluation.
All can be understood in intuitive terms, resulting from close-by or moderately
close neighbours.
Our results from the full simulation are consistent with the toy model calculation.
Though we have shown strong dependence on distance to the nearest neighbour 
(and thus on number density)
we found only weak sensitivity to neighbour brightness,
when averaged across broad bins of magnitude.
In addition to this, cuts on the DES Y1 catalogue sufficient to null the impact
of the detectable neighbours would result in a degradation of over 20\%
in source number density.
We cannot recommend such measures
for a code like \imshape,
in part because the data contains 
correlations between shear and 
number density.
Unless the link is preserved in the calibration simulations,
such selection could conceivably induce additional bias towards 
low shear\footnote{Although the sister catalogue to Y1 \imshape\
uses a form of internal calibration, which should allow one to correct for the 
additional selection bias.}.

Our investigation also assessed the impact of the faintest galaxies, which are not
reliably detected but nonetheless contribute flux to the survey images.
Via two different routes, first
using a spin-off neighbour-free resimulation,
and also using a subset of images simulated again with the sub-detection galaxies
missing, our findings suggested a net contribution to the multiplicative bias budget
of $m\sim-0.01$.

Unlike most earlier works on shear measurement,
we have propagated these findings to the most meaningful metric for cosmic
shear: 
bias on the inferred cosmological parameters.
The study we have presented here uses the DES Y1 cosmology pipeline,
as well as real non Gaussian shear covariance matrices
and photometric redshift distributions to implement MCMC forecasts.
In the first case considered, the data included a (different) multiplicative bias
in each redshift bin,
designed to approximate the residual $m$ that would arise were we to 
calibrate DES Y1 with a simple neighbour-free simulation.
Even marginalising over $m$ with a prior of $\mathcal{N}(0, 0.035)$
this scenario was demonstrated to result in a shift in the favoured cosmology towards
low clustering amplitude of more than $1\sigma$.

Finally, we have explored a second possible source of measurement bias arising from
the link between number density and neighbour bias. 
This enters two-point measurements as an additional correlation between the
multiplicative bias in galaxy pairs at small angular separation.
In the final section we have measured such a correlation from the \hoopoe\
mock images.
With the most pessimistic small-scale extrapolation, this was found to result in
a shift in the best-fitting cosmology of under $1\sigma$ in the negative $S_8$ direction,
which is not remedied by marginalising over $m$.
A less dramatic, though still considerable, increase in the correlation strength on small
scales was demonstrated to result in no discernable cosmological bias.

Both of these effects are of primary concern for the next generation of cosmological surveys.
By the end of their lifetime KiDS, DES and HSC are set to offer lensing-based cosmological constraints
comparable to the CMB.
The first, dominant, effect can be remedied relatively easily by
calibrating our shear measurements with sufficiently complex image simulations.
Indeed, the most recent shear constraints of 
\citet{hildebrandt16}, \citet{koehlinger17} and \citet{shearcorr}
have done just that.
Unfortunately, the correct treatment of scale dependent bias is not as clear,
though it should be captured at some level by the per-galaxy
responses upon which \metacal~relies.
Though further statements about the likely small scale dependence of the $mm$ correlation
are beyond the scope of the present study,
understanding this intricate topic will be crucial for future surveys if we are to
fully exploit the constraining power of the data.
The massive simulation efforts of LSST and Euclid,
combined with advancement in neighbour mitigation using techniques such as multi-object
fitting will be invaluable in this task.
With the enhanced understanding these will provide and the exquisite data of the next generation
surveys, the coming decade will be an exciting time for cosmology.

\section{Acknowledgements}

We thank Nicolas Tessore, Catherine Heymans and Rachel Mandelbaum for various insights that contributed to this work. We are also indebted to the many DES ``eyeballers" for lending their holiday time to help us understand and validate our simulations. The \hoopoe~simulations were generated using the National Energy Research Scientific Computing Center (NERSC) facility, which is maintained by the U.S. Department of Energy. The likelihood calculations were performed using NERSC and the Fornax computing cluster, which was funded by the European Research Council. 
SLB acknowledges support from the European Research Council in the form of a Consolidator Grant with number 681431.
Support for DG was provided by NASA through Einstein Postdoctoral Fellowship grant number PF5-160138 awarded by the Chandra X-ray Center, which is operated by the Smithsonian Astrophysical Observatory for NASA under contract NAS8-03060.

Funding for the DES Projects has been provided by the U.S. Department of Energy, the U.S. National Science Foundation, the Ministry of Science and Education of Spain, 
the Science and Technology Facilities Council of the United Kingdom, the Higher Education Funding Council for England, the National Center for Supercomputing 
Applications at the University of Illinois at Urbana-Champaign, the Kavli Institute of Cosmological Physics at the University of Chicago, 
the Center for Cosmology and Astro-Particle Physics at the Ohio State University,
the Mitchell Institute for Fundamental Physics and Astronomy at Texas A\&M University, Financiadora de Estudos e Projetos, 
Funda{\c c}{\~a}o Carlos Chagas Filho de Amparo {\`a} Pesquisa do Estado do Rio de Janeiro, Conselho Nacional de Desenvolvimento Cient{\'i}fico e Tecnol{\'o}gico and 
the Minist{\'e}rio da Ci{\^e}ncia, Tecnologia e Inova{\c c}{\~a}o, the Deutsche Forschungsgemeinschaft and the Collaborating Institutions in the Dark Energy Survey. 

The Collaborating Institutions are Argonne National Laboratory, the University of California at Santa Cruz, the University of Cambridge, Centro de Investigaciones Energ{\'e}ticas, 
Medioambientales y Tecnol{\'o}gicas-Madrid, the University of Chicago, University College London, the DES-Brazil Consortium, the University of Edinburgh, 
the Eidgen{\"o}ssische Technische Hochschule (ETH) Z{\"u}rich, 
Fermi National Accelerator Laboratory, the University of Illinois at Urbana-Champaign, the Institut de Ci{\`e}ncies de l'Espai (IEEC/CSIC), 
the Institut de F{\'i}sica d'Altes Energies, Lawrence Berkeley National Laboratory, the Ludwig-Maximilians Universit{\"a}t M{\"u}nchen and the associated Excellence Cluster Universe, 
the University of Michigan, the National Optical Astronomy Observatory, the University of Nottingham, The Ohio State University, the University of Pennsylvania, the University of Portsmouth, 
SLAC National Accelerator Laboratory, Stanford University, the University of Sussex, Texas A\&M University, and the OzDES Membership Consortium.

The DES data management system is supported by the National Science Foundation under Grant Numbers AST-1138766 and AST-1536171.
The DES participants from Spanish institutions are partially supported by MINECO under grants AYA2015-71825, ESP2015-88861, FPA2015-68048, SEV-2012-0234, SEV-2016-0597, and MDM-2015-0509, 
some of which include ERDF funds from the European Union. IFAE is partially funded by the CERCA program of the Generalitat de Catalunya.
Research leading to these results has received funding from the European Research
Council under the European Union's Seventh Framework Program (FP7/2007-2013) including ERC grant agreements 240672, 291329, and 306478.
We  acknowledge support from the Australian Research Council Centre of Excellence for All-sky Astrophysics (CAASTRO), through project number CE110001020.

This manuscript has been authored by Fermi Research Alliance, LLC under Contract No. DE-AC02-07CH11359 with the U.S. Department of Energy, Office of Science, Office of High Energy Physics. The United States Government retains and the publisher, by accepting the article for publication, acknowledges that the United States Government retains a non-exclusive, paid-up, irrevocable, world-wide license to publish or reproduce the published form of this manuscript, or allow others to do so, for United States Government purposes.

Based in part on observations at Cerro Tololo Inter-American Observatory, 
National Optical Astronomy Observatory, which is operated by the Association of 
Universities for Research in Astronomy (AURA) under a cooperative agreement with the National 
Science Foundation.

\bibliographystyle{mn2e}
\bibliography{samuroff,des_y1kp_short}

\appendix
\section{Derivation of a Two-Point Modifier for Scale Dependent Bias}\label{appendix:2pt_m}

In the following we set out a brief derivation of the analytic modifications to account for 
scale-dependent neighbour effects the shear-shear two-point correlations used in the earlier section. 
We do not claim that this is a precise calculation of the sort that could be used to derive a robust calibration. 
Rather it is an order of magnitude estimate to allow us to assess the approximate size of the cosmological bias 
these effects could induce in the data.

First, with complete generality it is possible to write the $i$ component of the measured shear at angular position $\theta$ as

\begin{equation}
\gamma_i^\mathrm{obs}(\boldsymbol{\theta}) = \left [ 1+m_i(\boldsymbol{\theta}) \right ] \gamma_i(\boldsymbol{\theta}),
\end{equation}

\noindent
where $\gamma_i$ is the underlying true shear, which is sensitive to cosmology only.
Extending this to the level of a two-point correlation
between two populations $\alpha$ and $\beta$ this implies:

\begin{multline}
\xi^{\mathrm{obs},\alpha\beta}_i (\theta) \equiv \left < \gamma^{\mathrm{obs},\alpha}_i(\boldsymbol{\theta}')\gamma^{\mathrm{obs},\beta}_i(\boldsymbol{\theta}'+\boldsymbol{\theta}) \right > _\theta \\
= \left < [1+m_i^\alpha(\boldsymbol{\theta}') ] [1+m_i^\beta(\boldsymbol{\theta}'+\boldsymbol{\theta}) ] \tilde{\gamma}_i^\alpha(\boldsymbol{\theta}') \tilde{\gamma}^\beta_i(\boldsymbol{\theta}'+\boldsymbol{\theta}) \right > _\theta.\\
\end{multline}

\noindent
Note that the observed shear used in a particular bin correlation is now weighted by the overdensity of galaxies in the image,
in addition to the calibration bias, such that

\begin{equation} 
\tilde{\gamma}_i^\alpha(\boldsymbol{\theta})\equiv \left [ 1+\delta_g^\alpha(\boldsymbol{\theta}) \right] \times \gamma_i^\alpha(\boldsymbol{\theta}).
\end{equation}

\noindent
Expanding each of the terms one finds:

\begin{multline}
\xi^{\mathrm{obs},\alpha\beta}_i (\theta) = \left < \gamma_i^\alpha(\boldsymbol{\theta}') \gamma_i^\beta (\boldsymbol{\theta}'+ \boldsymbol{\theta}) \right >_\theta\\ 
+ \left < m^\alpha_i(\boldsymbol{\theta}')  \gamma_i^\alpha(\boldsymbol{\theta}')  \gamma_i^\beta (\boldsymbol{\theta}'+ \boldsymbol{\theta}) \right >_\theta 
+ \left < m^\beta_i(\boldsymbol{\theta}'+\boldsymbol{\theta}) \gamma^\alpha_i(\boldsymbol{\theta}') \gamma^\beta_i(\boldsymbol{\theta}'+\boldsymbol{\theta}) \right >_\theta \\
+ \left < \delta_g^\alpha(\boldsymbol{\theta}')  \gamma_i^\alpha(\boldsymbol{\theta}')  \gamma_i^\beta (\boldsymbol{\theta}'+ \boldsymbol{\theta}) \right >_\theta 
+ \left < \delta_g^\beta(\boldsymbol{\theta}'+\boldsymbol{\theta}) \gamma^\alpha_i(\boldsymbol{\theta}') \gamma^\beta_i(\boldsymbol{\theta}'+\boldsymbol{\theta}) \right >_\theta \\
+ \left < m^\alpha_i(\boldsymbol{\theta}') m^\beta_i(\boldsymbol{\theta}'+\boldsymbol{\theta}) \gamma^\alpha_i(\boldsymbol{\theta}') \gamma^\beta_i(\boldsymbol{\theta}'+\boldsymbol{\theta}) \right >_\theta\\
+ \left < \delta_g^\alpha(\boldsymbol{\theta}') \delta_g^\beta(\boldsymbol{\theta}'+\boldsymbol{\theta}) \gamma^\alpha_i(\boldsymbol{\theta}') \gamma^\beta_i(\boldsymbol{\theta}'+\boldsymbol{\theta}) \right >_\theta.\\
\end{multline}

\noindent
The terms contributing to the measured two-point shear correlation, then, is sensitive to both
spatial correlations between the $m$ in different galaxies
and to the correlations with the source density.
Note that we've chosen to neglect a higher-order (six-point) term.
In reality there will also be a connection between galaxy density and shear,
but we will follow the normal convention and assume the contribution is small enough to be
neglected. 
In simple terms, an excess in the $\left < mm \right >$ term above the product of the
mean $m$ values indpendently could arise because galaxy pairs separated on small scales
tend to come from \emph{similar} image plane environments.
In contrast the density weighted correlations $\left < \delta_g m \right >$ would be zero,
but for a simple observation; 
selecting a random galaxy with a suitable correlation pair at a distance $\theta$
is \emph{not} the same as unconditionally selecting a random galaxy.
In the small scale bins we will over-sample the dense 
regions, where $m$ tends to be larger
(see Section \ref{subsection:cosmology:scale_dependent_m}).
The angular brackets here indicate averaging over all galaxy pairs separated by $\theta$.
If we can assume the bias is independent of the underlying cosmology the above expression 
simplifies significantly:

\begin{multline}\label{eq:xi_obs}
\xi^{\mathrm{obs},\alpha\beta}_i (\theta) =  ( 1 + \bar{m}_i^\alpha  + \bar{m}_i^\beta   + \left < m^\alpha_i(\boldsymbol{\theta}') m^\beta_i (\boldsymbol{\theta}'+\boldsymbol{\theta}) \right > _\theta \\
+ \left < \delta^\alpha_g(\boldsymbol{\theta}') m^\beta_i (\boldsymbol{\theta}'+\boldsymbol{\theta}) \right > _\theta 
+ \left < m^\alpha_i(\boldsymbol{\theta}') \delta^\beta_g (\boldsymbol{\theta}'+\boldsymbol{\theta}) \right > _\theta \\
+ \left < \delta^\alpha_g(\boldsymbol{\theta}') \delta^\beta_g (\boldsymbol{\theta}'+\boldsymbol{\theta}) \right > _\theta )
\times \xi^{\alpha\beta}_i (\theta | \textbf{p} ),
\end{multline}

\noindent
with $\xi_i^{\alpha\beta}$ being the true correlation function of cosmological shears $\left < \gamma_i \gamma_i \right >$, 
which is contingent on the underlying cosmological parameters \textbf{p}.
It can be shown that 

\begin{multline}
\xi_+(\theta) \equiv \left < \gamma_+(\boldsymbol{\theta}') \gamma_+(\boldsymbol{\theta}'+\boldsymbol{\theta}) \right >_\theta 
\pm \left < \gamma_\times(\boldsymbol{\theta}') \gamma_\times(\boldsymbol{\theta}'+\boldsymbol{\theta}) \right >_\theta\\
=\left < \gamma_1(\boldsymbol{\theta}') \gamma_1(\boldsymbol{\theta}'+\boldsymbol{\theta}) \right >_\theta 
\pm \left < \gamma_2(\boldsymbol{\theta}') \gamma_2(\boldsymbol{\theta}'+\boldsymbol{\theta}) \right >_\theta\\
= \xi_1 (\theta) + \xi_2 (\theta),
\end{multline}

\noindent
and so one can use equation \ref{eq:xi_obs} to construct the observed $\xi_\pm$ correlation functions

\begin{multline}\label{eq:all_prefactors}
\xi^{\mathrm{obs},\alpha\beta}_\pm (\theta) = \biggl ( 1 + \bar{m}^\alpha + \bar{m}^\beta
+ \left < m^\alpha(\boldsymbol{\theta}') m^\beta (\boldsymbol{\theta}'+\boldsymbol{\theta}) \right > _\theta \\
+ \left < \delta_g^\alpha(\boldsymbol{\theta}') m^\beta (\boldsymbol{\theta}'+\boldsymbol{\theta}) \right > _\theta 
+ \left < m^\alpha(\boldsymbol{\theta}') \delta_g^\beta (\boldsymbol{\theta}'+\boldsymbol{\theta}) \right > _\theta \\
+ \left < \delta_g^\alpha(\boldsymbol{\theta}') \delta_g^\beta (\boldsymbol{\theta}'+\boldsymbol{\theta}) \right > _\theta 
\biggr ) \xi^{\alpha\beta}_\pm (\theta | \textbf{p} ).
\end{multline}

\noindent
The $i$ subscript has been discarded here under the assumption that $m_1$ and $m_2$ are approximately equal
for a given set of galaxies.

Next, let's say imagine that we have a measured datavector. Our measurements are biased,
but we'll assume it is possible to devise a correction that recovers the true cosmological signal precisely.
Our observed datavector is then just,

\begin{equation}\label{eq:xi_be}
\xi^{\mathrm{obs},\alpha\beta}_\pm (\theta) = \Upsilon^{\mathrm{tr},\alpha\beta} \xi^{\alpha\beta}_\pm (\theta | \textbf{p} ), \\
\end{equation}

\noindent
which follows trivially from equation \ref{eq:all_prefactors}.
Since we do not trivially know $\Upsilon^{\mathrm{tr},\alpha\beta}$ \emph{ab initio}
(that's why we need simulations!)
we can only construct a best-estimate approximation.
By applying a correction factor to the raw measurements we construct a best-estimate datavector:

\begin{equation}\label{eq:xi_rescaling_tr}
\xi^{\mathrm{BE},\alpha\beta}_\pm (\theta) = \frac{1}{\Upsilon^{\mathrm{BE},\alpha\beta}} \xi^{\mathrm{obs},\alpha\beta}_\pm (\theta)
 = \frac{\Upsilon^{\mathrm{tr},\alpha\beta}} {\Upsilon^{\mathrm{BE},\alpha\beta}} \xi^{\alpha\beta}_\pm (\theta | \textbf{p} ).
\end{equation} 

\noindent
Of course, if our best correction is perfect then the ratio goes to unity,
and we recover the underlying cosmology.
Since we apply corrections to the single-galaxy shears we will assume $\Upsilon^{\mathrm{BE},\alpha\beta}$
includes the $\left < \delta_g\delta_g \right>$ term, but neglects the correlations involving $m$.
We then can write:

\begin{equation}
\Upsilon^{\mathrm{BE},\alpha\beta}
= \left ( 1 + \bar{m}^{\alpha} + \bar{m}^{\beta}  + \bar{m}^{\alpha}\bar{m}^{\beta} 
+ \left < \delta_g^\alpha(\boldsymbol{\theta}') \delta_g^\beta (\boldsymbol{\theta}'+\boldsymbol{\theta}) \right > _\theta \right ).
\end{equation}

\noindent
We can measure the mean bias in each bin that would be obtained from the calibration directly.
As we show in \citetalias{shearcat}, using the full DES Y1 \hoopoe~ catalogues, these biases are $\sim -0.08$ to $-0.20$.

Finally, assume that although $m$ clearly varies between redhshift bins,
the strength of the correlation does not 
That is, the bias-bias term is the product of the mean $m$s
(which varies between $z$ bins)
plus a scale dependent shift (which doesn't).
One then has:

\begin{equation}
\left < m^{\alpha}(\boldsymbol{\theta}') m^{\beta}(\boldsymbol{\theta}'+\boldsymbol{\theta}) \right > _\theta
= \bar{m}^{\alpha} \bar{m}^{\beta} + \Delta \xi_{mm}(\boldsymbol{\theta}).
\end{equation}

\noindent
The additive part can be measured directly from the simulation using sub-patches, as described earlier.
The density-density correlation can be obtained in the same way.
This, then, leaves only the $m \times \delta_g$ cross-correlation.
This should vanish in the case of zero correlation, but it also seems reasonable to assume that
the magnitude should be proportional to the mean bias $\bar{m}^{\alpha}$ in a particular bin.
This allows the scale dependent (non-tomographic) cross correlation measured from \hoopoe\
to be rescaled appropriately for each bin pair:

 \begin{equation}
\left < \delta^{\alpha}(\boldsymbol{\theta}') m^{\beta}(\boldsymbol{\theta}'+\boldsymbol{\theta}) \right > _\theta
= \left ( \frac{\bar{m}^{\beta}}{\bar{m}} \right ) \xi_{gm}(\boldsymbol{\theta}),
\end{equation}

\noindent
where $\bar{m}$ is the global multiplicative bias and $\xi_{gm}(\boldsymbol{\theta}) \equiv \left< m \delta_g \right >$,
each measured using all simulated galaxies.
Using the above equations, with our fiducial calibration and three measured correlations,
one can derive a scale dependent modification to shear-shear two point correlation data using equation \ref{eq:xi_be}.

\section*{Affiliations}
\input{affiliations}

\label{lastpage}

\end{document}

%% file: authors.tex
\author[S. Samuroff, S.~L. Bridle, J. Zuntz et al]{
\parbox{\textwidth}{
\Large
S.~Samuroff$^{1}$\thanks{\texttt{simon.samuroff@postgrad.manchester.ac.uk}},
S.~L.~Bridle$^{1}$,
J.~Zuntz$^{2}$,
M.~A.~Troxel$^{3,4}$,
D.~Gruen$^{5,6}$\thanks{NASA Einstein Fellow},
R.~P.~Rollins$^{1}$,
G.~M.~Bernstein$^{7}$,
T.~F.~Eifler$^{8}$,
E.~M.~Huff$^{8}$,
T.~Kacprzak$^{9}$,
E.~Krause$^{5}$,
N.~MacCrann$^{3,4}$,
F.~B.~Abdalla$^{10,11}$,
S.~Allam$^{12}$,
J.~Annis$^{12}$,
K.~Bechtol$^{13}$,
A.~Benoit-L{\'e}vy$^{14,10,15}$,
E.~Bertin$^{14,15}$,
D.~Brooks$^{10}$,
E.~Buckley-Geer$^{12}$,
A. Carnero Rosell$^{16,17}$,
M.~Carrasco~Kind$^{18,19}$,
J.~Carretero$^{20}$,
M.~Crocce$^{21}$,
C.~B.~D'Andrea$^{7}$,
L.~N.~da Costa$^{16,17}$,
C.~Davis$^{5}$,
S.~Desai$^{22}$,
P.~Doel$^{10}$,
A.~Fausti Neto$^{16}$,
B.~Flaugher$^{12}$,
P.~Fosalba$^{21}$,
J.~Frieman$^{12,23}$,
J.~Garc\'ia-Bellido$^{24}$,
D.~W.~Gerdes$^{25,26}$,
R.~A.~Gruendl$^{18,19}$,
J.~Gschwend$^{16,17}$,
G.~Gutierrez$^{12}$,
K.~Honscheid$^{3,4}$,
D.~J.~James$^{27,28}$,
M.~Jarvis$^{7}$,
T.~Jeltema$^{29}$,
D.~Kirk$^{10}$,
K.~Kuehn$^{30}$,
S.~Kuhlmann$^{31}$,
T.~S.~Li$^{12}$,
M.~Lima$^{32,16}$,
M.~A.~G.~Maia$^{16,17}$,
M.~March$^{7}$,
J.~L.~Marshall$^{33}$,
P.~Martini$^{3,34}$,
P.~Melchior$^{35}$,
F.~Menanteau$^{18,19}$,
R.~Miquel$^{36,20}$,
B.~Nord$^{12}$,
R.~L.~C.~Ogando$^{16,17}$,
A.~A.~Plazas$^{8}$,
A.~Roodman$^{5,6}$,
E.~Sanchez$^{37}$,
V.~Scarpine$^{12}$,
R.~Schindler$^{6}$,
M.~Schubnell$^{26}$,
I.~Sevilla-Noarbe$^{37}$,
E.~Sheldon$^{38}$,
M.~Smith$^{39}$,
M.~Soares-Santos$^{12}$,
F.~Sobreira$^{40,16}$,
E.~Suchyta$^{41}$,
G.~Tarle$^{26}$,
D.~Thomas$^{42}$,
D.~L.~Tucker$^{12}$
\begin{center} (DES Collaboration) \end{center}
}
\vspace{0.4cm}
\\
\parbox{\textwidth}{Author affiliations are listed at the end of the paper.}
}

%% file: affiliations.tex
$^{1}$ Jodrell Bank Centre for Astrophysics, School of Physics and Astronomy, University of Manchester, Oxford Road, Manchester, M13 9PL, UK\\
$^{2}$ Institute for Astronomy, University of Edinburgh, Edinburgh EH9 3HJ, UK\\
$^{3}$ Center for Cosmology and Astro-Particle Physics, The Ohio State University, Columbus, OH 43210, USA\\
$^{4}$ Department of Physics, The Ohio State University, Columbus, OH 43210, USA\\
$^{5}$ Kavli Institute for Particle Astrophysics \& Cosmology, P. O. Box 2450, Stanford University, Stanford, CA 94305, USA\\
$^{6}$ SLAC National Accelerator Laboratory, Menlo Park, CA 94025, USA\\
$^{7}$ Department of Physics and Astronomy, University of Pennsylvania, Philadelphia, PA 19104, USA\\
$^{8}$ Jet Propulsion Laboratory, California Institute of Technology, 4800 Oak Grove Dr., Pasadena, CA 91109, USA\\
$^{9}$ Department of Physics, ETH Z\"urich, Wolfgang-Pauli-Strasse 16, CH-8093 Z\"urich, Switzerland\\
$^{10}$ Department of Physics \& Astronomy, University College London, Gower Street, London, WC1E 6BT, UK\\
$^{11}$ Department of Physics and Electronics, Rhodes University, PO Box 94, Grahamstown, 6140, South Africa\\
$^{12}$ Fermi National Accelerator Laboratory, P. O. Box 500, Batavia, IL 60510, USA\\
$^{13}$ LSST, 933 North Cherry Avenue, Tucson, AZ 85721, USA\\
$^{14}$ CNRS, UMR 7095, Institut d'Astrophysique de Paris, F-75014, Paris, France\\
$^{15}$ Sorbonne Universit\'es, UPMC Univ Paris 06, UMR 7095, Institut d'Astrophysique de Paris, F-75014, Paris, France\\
$^{16}$ Laborat\'orio Interinstitucional de e-Astronomia - LIneA, Rua Gal. Jos\'e Cristino 77, Rio de Janeiro, RJ - 20921-400, Brazil\\
$^{17}$ Observat\'orio Nacional, Rua Gal. Jos\'e Cristino 77, Rio de Janeiro, RJ - 20921-400, Brazil\\
$^{18}$ Department of Astronomy, University of Illinois, 1002 W. Green Street, Urbana, IL 61801, USA\\
$^{19}$ National Center for Supercomputing Applications, 1205 West Clark St., Urbana, IL 61801, USA\\
$^{20}$ Institut de F\'{\i}sica d'Altes Energies (IFAE), The Barcelona Institute of Science and Technology, Campus UAB, 08193 Bellaterra (Barcelona) Spain\\
$^{21}$ Institute of Space Sciences, IEEC-CSIC, Campus UAB, Carrer de Can Magrans, s/n,  08193 Barcelona, Spain\\
$^{22}$ Department of Physics, IIT Hyderabad, Kandi, Telangana 502285, India\\
$^{23}$ Kavli Institute for Cosmological Physics, University of Chicago, Chicago, IL 60637, USA\\
$^{24}$ Instituto de Fisica Teorica UAM/CSIC, Universidad Autonoma de Madrid, 28049 Madrid, Spain\\
$^{25}$ Department of Astronomy, University of Michigan, Ann Arbor, MI 48109, USA\\
$^{26}$ Department of Physics, University of Michigan, Ann Arbor, MI 48109, USA\\
$^{27}$ Astronomy Department, University of Washington, Box 351580, Seattle, WA 98195, USA\\
$^{28}$ Cerro Tololo Inter-American Observatory, National Optical Astronomy Observatory, Casilla 603, La Serena, Chile\\
$^{29}$ Santa Cruz Institute for Particle Physics, Santa Cruz, CA 95064, USA\\
$^{30}$ Australian Astronomical Observatory, North Ryde, NSW 2113, Australia\\
$^{31}$ Argonne National Laboratory, 9700 South Cass Avenue, Lemont, IL 60439, USA\\
$^{32}$ Departamento de F\'isica Matem\'atica, Instituto de F\'isica, Universidade de S\~ao Paulo, CP 66318, S\~ao Paulo, SP, 05314-970, Brazil\\
Station, TX 77843,  USA\\
$^{34}$ Department of Astronomy, The Ohio State University, Columbus, OH 43210, USA\\
$^{35}$ Department of Astrophysical Sciences, Princeton University, Peyton Hall, Princeton, NJ 08544, USA\\
$^{36}$ Instituci\'o Catalana de Recerca i Estudis Avan\c{c}ats, E-08010 Barcelona, Spain\\
$^{37}$ Centro de Investigaciones Energ\'eticas, Medioambientales y Tecnol\'ogicas (CIEMAT), Madrid, Spain\\
$^{38}$ Brookhaven National Laboratory, Bldg 510, Upton, NY 11973, USA\\
$^{39}$ School of Physics and Astronomy, University of Southampton,  Southampton, SO17 1BJ, UK\\
$^{40}$ Instituto de F\'isica Gleb Wataghin, Universidade Estadual de Campinas, 13083-859, Campinas, SP, Brazil\\
$^{41}$ Computer Science and Mathematics Division, Oak Ridge National Laboratory, Oak Ridge, TN 37831\\
$^{42}$ Institute of Cosmology \& Gravitation, University of Portsmouth, Portsmouth, PO1 3FX, UK\\